# Symmetries and selection rules in Floquet systems: application to harmonic generation in nonlinear optics

*Ofer Neufeld[1,2], Daniel Podolsky[2] and Oren Cohen[1,2]*

[1]Solid state Institute, Technion - Israel Institute of Technology, Haifa 32000, Israel.

[2]Physics Department, Technion - Israel Institute of Technology, Haifa 32000, Israel.

Corresponding Authors e-mails: ofern@tx.technion.ac.il, oren@tx.technion.ac.il

Symmetry is one of the most generic and useful concepts in physics and chemistry, often leading to conservation laws and selection rules. For example, symmetry considerations have been used to predict selection rules for transitions in atoms, molecules, and solids. Floquet systems also demonstrate a variety of symmetries which are spatiotemporal (i.e. dynamical symmetries (DSs)). However, the derivation of selection rules from DSs has so far been limited to several ad hoc cases. A general theory for deducing the impact of DSs in physical systems has not been formulated yet. Here we explore symmetries exhibited in Floquet systems using group theory, and discover novel DSs and selection rules. We derive the constraints on a general system's temporal evolution, and selection rules that are imposed by the DSs. As an example, we apply the theory to harmonic generation, and derive tables linking (2+1)D and (3+1)D DSs of the driving laser and medium to allowed harmonic emission and its polarization. We identify several new symmetries and selection rules, including an elliptical DS that leads to production of elliptically polarized harmonics where all the harmonics have the same ellipticity, and selection rules that have no explanation based on currently known conservation laws. We expect the theory to be useful for manipulating the harmonic spectrum, and for ultrafast spectroscopy. Furthermore, the presented Floquet group theory should be useful in various other systems, e.g., Floquet topological insulators and photonic lattices, possibly yielding formal and general classification of symmetry and topological properties.

Symmetry has been used as a principle concept throughout science[1,2]. It simplifies problems and hints to initial ansatz solutions. For example, symmetries give rise to conserved quantities, and to selection rules for electronic transitions in molecules and solids[3,4]. Time-periodic Floquet systems exhibit spatiotemporal symmetries (denoted dynamical symmetries – DSs). Some examples include perturbative[5–7] and high harmonic generation (HHG)[8,9], Floquet topological insulators[10–13], Floquet-Weyl semimetals[14,15], photonic lattices (that are modulated along the propagation axis)[16,17], solid state lattices[10,18–21], and Bose-Einstein condensates[22–24]. While some selection rules were derived from DSs[25–27], a general theory for the manifestation of DSs in physical systems has not been formulated yet.

Mathematically, symmetries are best described by group theory, which provides a rigorous platform for their analysis. Space-time groups that describe DSs (excluding time-reversal symmetries) were presented in the 60's[28], yet they were not utilized for exploring Floquet systems. A closely related theory was developed in the 70's describing so-called 'line-groups'[29,30], which characterize the spatial symmetries of polymers. The selection rules in polymers are determined by analyzing the system's eigenstates and transition probabilities under finite perturbations. This approach however, is inappropriate for Floquet systems because they are continuously excited by a strong time-dependent perturbation that requires a dynamical theory. Selection rules in Floquet systems arise from the actual time-dependent dynamics, which is generally unknown.

Here, we use a general group theory approach to explore symmetries and selection rules in Floquet systems. We apply the theory to harmonic generation (HG), including the pertubative[31], and non-perturbative HHG regime[8,9], which is used as a unique table-top source of coherent radiation in Extreme UV and X-ray spectral regions, and for producing attosecond pulses for ultrafast spectroscopy[9]. HG is greatly affected by DSs in the driving laser and material target that dictate the allowed emission. For instance, HG by a half-wave symmetric driving laser that interacts with isotropic media results in odd-only harmonics[25]. Additionally, discrete rotational DSs have been employed for generating circularly polarized high harmonics using bi-circular two-color laser fields[32–40], as well as other combinations of drivers and molecular targets[26,27,41–45]. Notably, all known selection rules in HG can be explained and derived by DSs, or equivalently by conservation laws.

Here we explore the symmetries in Floquet systems using group theory. We systematically derive DSs as generalized products of spatial and temporal transformations in both (2+1)D and (3+1)D, yielding 'dynamical groups' that describe the symmetries of time-periodic systems. We prove that if a time-dependent Hamiltonian commutes with a 'dynamical group', then the group's generating operators dictate the evolution of all the physical observables in the system, and we derive the resulting restrictions for all DSs. We apply the theory to HG, introducing several novel DSs that lead to new selection rules, including an elliptical symmetry that can be used to control the polarization properties of high harmonics. Remarkably, some of the new selection rules are not explained by currently known conservation laws.

**Symmetry elements in Floquet systems**

The symmetries of Floquet systems can be described by adjoining point group-like dimensions (i.e., molecular) that describe spatial symmetries, and an infinite and periodic dimension which is space group-like (i.e., lattice-like) that describes the $T$-periodic time axis and temporal symmetries. For simplicity, we exclude spatial translational symmetries (extension to time-crystals are possible), and only deal with vectorial time-dependent functions (though the method also applies to scalar functions). Considering a general time dependent vector field $\vec{E}(t)$, a symmetry of the field is an operation that leaves it invariant. Thus $\hat{X}$ is a symmetry of $\vec{E}(t)$ if $\hat{X} \cdot \vec{E}(t) = \vec{E}(t)$. $\hat{X}$ can be a purely spatial operation, a purely temporal operation, or a product of temporal and spatial operations. In the (2+1)D case, there are only two relevant types of spatial symmetry elements: rotations, denoted by the operator $\hat{R}_n$ ($\hat{R}_n$ stands for rotation by an angle $2\pi/n$), and reflections, denoted by the operator $\hat{\sigma}$. In (3+1)D we also consider spatial inversion $\hat{\imath}$, and improper rotations $\hat{s}_n = \hat{\sigma}_h \cdot \hat{R}_n$ ($\hat{\sigma}_h$ stands for reflection in the plane perpendicular to the adjoined rotation axis). Temporal symmetry elements include time-reversal, denoted by $\hat{T}$ (which is order 2), and time-translations, where translations by time $T/n$ are denoted by $\hat{\tau}_n$ (which is order $n$).

We derive here a general theory for Floquet systems by considering all products of spatial and temporal operations that close under group multiplication. We exclude purely spatial transformations[4] since they arise only in trivial cases (e.g., reduced dimensionality) where the symmetry group is a direct product of the spatial and dynamical groups. This exclusion greatly reduces the amount of DSs that permit closure. For instance, consider a general DS with temporal and spatial parts, $\hat{X}_t \cdot \hat{X}_s$. If this DS is raised to the order of $\hat{X}_t$, then $\hat{X}_s$ raised to the order of $\hat{X}_t$ must be the spatial identity (otherwise, we get a purely spatial transformation). This, combined with the fact that T is the basic period of $\vec{E}(t)$, implies that either $\hat{X}_t$ and $\hat{X}_s$ have the same order, or $\hat{X}_s$ is the spatial identity, permitting purely temporal DSs. In what follows, this approach is used to systematically derive all DSs in Floquet groups according to their operation order.

## Dynamical symmetries and groups in (2+1)D

We start out by considering the (2+1)D case, and derive DSs that are products of 2D spatial and temporal operations. Adjoining order 2 temporal operators with spatial reflection yields the following DSs (Figure 1):

$$\hat{D} = \hat{T} \cdot \hat{\sigma} \tag{1}$$
$$\hat{Z} = \hat{\tau}_2 \cdot \hat{\sigma} \tag{2}$$
$$\hat{H} = \hat{T} \cdot \hat{\tau}_2 \cdot \hat{\sigma} \tag{3}$$

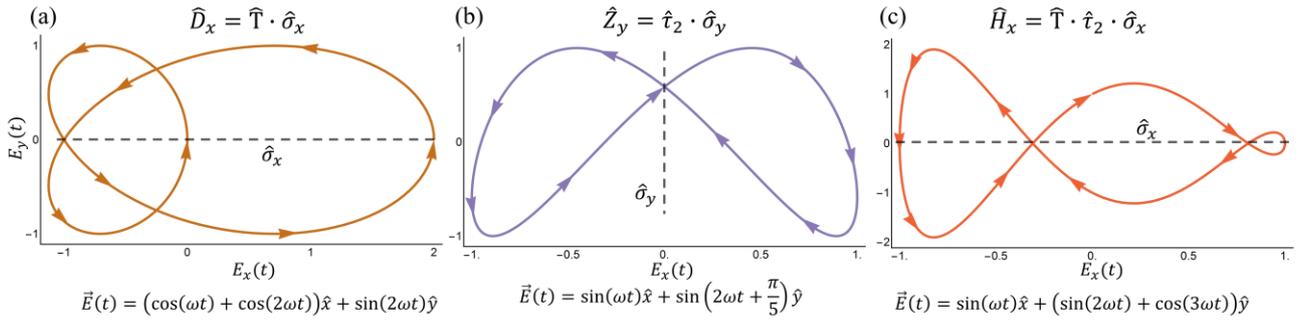

**Figure 1| Order-2 spatiotemporal DSs involving spatial reflection with examples for time-periodic fields possessing each symmetry. a** $\hat{D}_x$ symmetry, **b** $\hat{Z}_y$ symmetry, and **c** $\hat{H}_x$ symmetry. The fields are represented on Lissajou plots. The spatial part of the operators is indicated by dashed lines, colored arrows along the plot indicate the direction of time.

The same temporal operations can also be adjoined to rotations by 180º (Figure 2):

$$\hat{C}_2 = \hat{\tau}_2 \cdot \hat{R}_2 \tag{4}$$
$$\hat{Q} = \hat{T} \cdot \hat{R}_2 \tag{5}$$
$$\hat{G} = \hat{T} \cdot \hat{\tau}_2 \cdot \hat{R}_2 \tag{6}$$

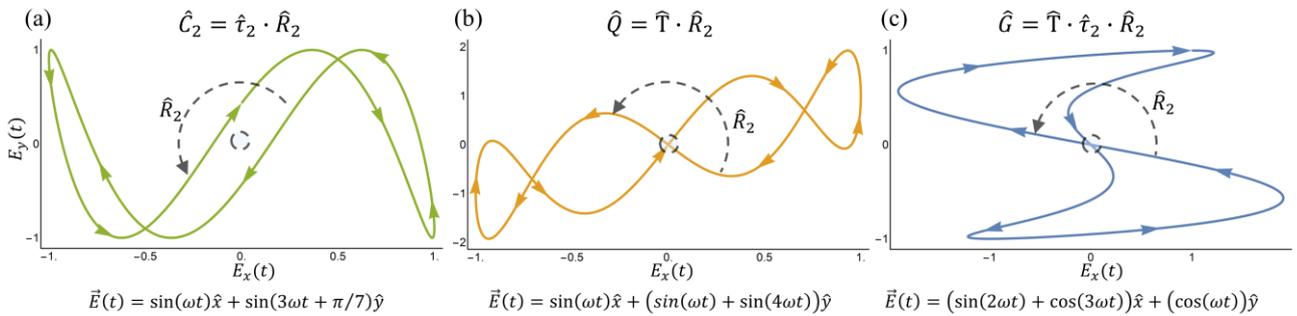

**Figure 2| Order-2 spatiotemporal DSs involving spatial rotations by 180º with examples for time-periodic fields possessing each symmetry. a** $\hat{C}_2$ symmetry, **b** $\hat{Q}$ symmetry, and **c** $\hat{G}$ symmetry. The fields are represented on Lissajou plots. The spatial part of the operators is indicated by dashed arrows, colored arrows along the plot indicate the direction of time.

Next, we map out higher order operators that involve rotations (temporal screw-axes):

$$\hat{C}_n = \hat{\tau}_n \cdot \hat{R}_n \tag{7}$$

The operator $\hat{C}_n$ is a DS of bi-circular EM fields, and of the Hamiltonians of circularly polarized electric fields interacting with rotationally invariant molecules[26,35,36,38,39,45] (Figure 3a). The $\hat{C}_n$ operator is generalized to cases where the spatial rotations return to the identity after more than one cycle (Figure 3b), denoted $\hat{C}_{n,m}$. This type of DS is exhibited by bi-circular EM fields of frequencies $m\omega$ and $(n-m)\omega$[38,46], and is expressed as:

$$\hat{C}_{n,m} = \hat{\tau}_n \cdot (\hat{R}_n)^m \equiv \hat{\tau}_n \cdot \hat{R}_{n,m} \tag{8}$$

For $m=1$ this operator reduces to Eq. (7).

The above DSs contain spatial transformations that are all symmetries in molecular groups[4]. We have also discovered a new type of DS with a spatial term that has no analogue in molecules. This is a discrete elliptical symmetry that generalizes Eq. (8) by considering rotations along an ellipse instead of a circle. By convention, we define elliptical symmetries to always have their major axis along the $x$-axis. This symmetry is expressed as products of rotation and scaling operators:

$$\hat{e}_{n,m} = \hat{\tau}_n \cdot \hat{L}_b \cdot \hat{R}_{n,m} \cdot \hat{L}_{1/b} \text{ , where} \tag{9}$$

$$\hat{L}_b = \begin{pmatrix} 1 & \\ & b \end{pmatrix} \tag{10}$$

is a scaling transformation along the elliptical axis spanned in Cartesian coordinates, and $0 \leq b \leq 1$ is the ellipticity of the underlying symmetry (see example 3$^{rd}$ order case in Figure 3c).

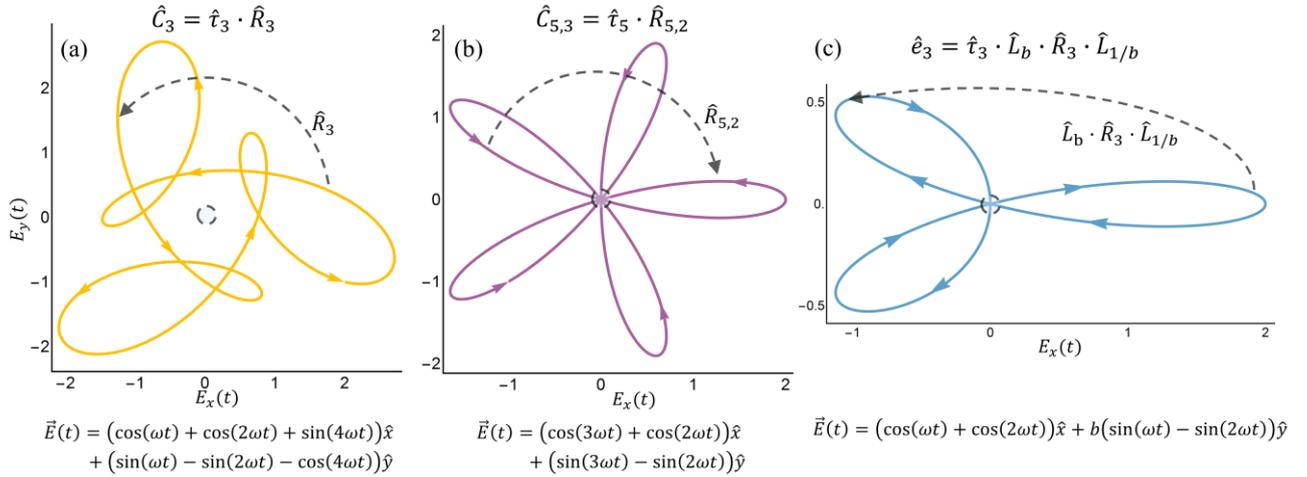

**Figure 3| High order spatiotemporal DSs with examples of time-periodic fields possessing each symmetry. a** $\hat{C}_3$ symmetry, **b** $\hat{C}_{5,3}$ symmetry, and **c** $\hat{e}_3$ symmetry. The fields are represented on Lissajou plots. The spatial part of the operators is indicated by dashed arrows, colored arrows along the plot indicate the direction of time.

The DSs described above form the basis of 'dynamical groups'. Physically, this means that a specific Floquet system can simultaneously exhibit several types of DSs, that is, its symmetry properties are uniquely defined by a set of generating operators. We mention several examples (more examples are given in section S.4 in the supplementary information (SI)). First, there are order 2 groups with one generator (examples seen in Figure 1 and 2), and cyclic groups of higher order that have $\hat{C}_{n,m}$ or $\hat{e}_{n,m}$ as generators (Figure 3a), which are all abelian. Groups with two generators can be either abelian (Figure 4a and 4b), or non-abelian (Figure 4c).

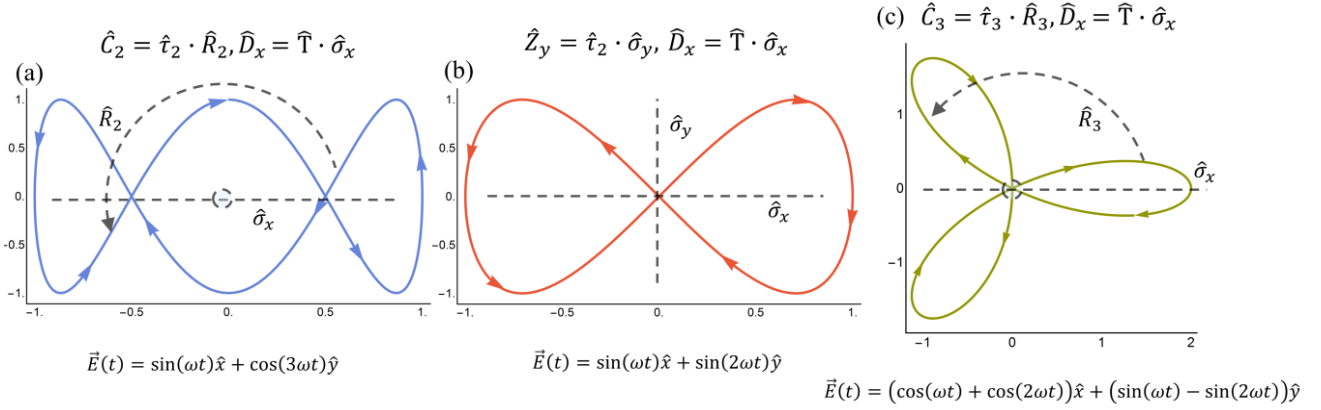

**Figure 4| Example time-periodic fields characterized by groups with two generators. a** A group with the generators $\hat{C}_2$ and $\widehat{D}_x$, **b** a group with the generators $\hat{Z}_y$ and $\widehat{D}_x$, and **c** a group with the generators $\hat{C}_3$ and $\widehat{D}_x$. The fields are represented on Lissajou plots. The spatial parts of the operators are indicated by dashed lines and arrows, colored arrows along the plot indicate the direction of time.

## Dynamical symmetries and groups in (3+1)D

We now consider the full (3+1)D case. In terms of the symmetry operations themselves, this means upgrading reflection axes to planes, and rotational operators to revolve around a specific axis. Thus, all DSs in Eqs. (1)-(10) are extended to the 3D case by choice of plane or axis. For example, elliptical rotations are extended to 3D by denoting the $z$-axis as the rotation axis, and specifying the axis around which the scaling transformation occurs (within the $xy$ plane):

$$\hat{e}_{n,m} = \hat{\tau}_n \cdot \hat{L}^y{}_b \cdot \hat{R}^z{}_{n,m} \cdot \hat{L}^y{}_{1/b}, \text{ where} \tag{11}$$

$$\hat{L}^y{}_b = \begin{pmatrix} 1 & & \\ & b & \\ & & 1 \end{pmatrix} \tag{12}$$

An example for the (3+1)D case of the $\hat{C}_2$ DS is shown in Figure 5(a). Beyond these DSs, (3+1)D also offers completely new DSs that have no analogue in (2+1)D. These include order 2 DSs with spatial inversion appended to temporal operations:

$$\hat{J} = \widehat{T} \cdot \hat{\imath} \tag{13}$$
$$\hat{F} = \hat{\tau}_2 \cdot \hat{\imath} \tag{14}$$
$$\hat{A} = \widehat{T} \cdot \hat{\tau}_2 \cdot \hat{\imath} \tag{15}$$

, and DSs that involve spatial improper rotations, which are products of a rotation and a reflection about the plane normal to the rotation axis. There are two different improper rotational DSs, for even or odd orders, respectively. For even orders we define the DS:

$$\widehat{M}_{2n,m} = \hat{\tau}_{2n} \cdot \hat{\sigma}_h \cdot \hat{R}_{2n,m} \equiv \hat{\tau}_{2n} \cdot \hat{s}_{2n,m}, \text{ whereas for odd orders we have:} \tag{16}$$
$$\widehat{M}_{2n+1,m} = \hat{\tau}_{2(2n+1)} \cdot \hat{s}_{2n+1,m} \tag{17}$$

where $\widehat{M}_{2n,m}$ is order $2n$, and $\widehat{M}_{2n+1,m}$ is order $2(2n+1)$, and the index $m$ is equivalent to that in Eq. (8) (for example see Figure 5(c)). The operator in Eq. (16) was previously considered in cross-beam geometries[27]. Naturally, these symmetries can be generalized to an elliptical case by replacing the standard circular rotations with elliptical ones. This leads to improper elliptical rotation DSs:

$$\hat{P}_{2n,m} = \hat{\tau}_{2n} \cdot \hat{\sigma}_h \cdot (\hat{L}_b \cdot \hat{R}_{2n,m} \cdot \hat{L}_{1/b}) = \hat{e}_{2n,m} \cdot \hat{\sigma}_h \tag{18}$$
$$\hat{P}_{2n+1,m} = \hat{\tau}_{2(2n+1)} \cdot \hat{\sigma}_h \cdot (\hat{L}_b \cdot \hat{R}_{2n+1,m} \cdot \hat{L}_{1/b}) \tag{19}$$

All of the (3+1)D DSs construct similar groups to the (2+1)D case that have three finite spatial dimensions, and an additional infinite and periodic time axis. For example, the field shown in Figure 5(c) possesses not only improper rotational DS of order 4, but also a 2-fold rotational DS around the z-axis ($\hat{C}_2$), as well as other DSs which all together form a dynamical group.

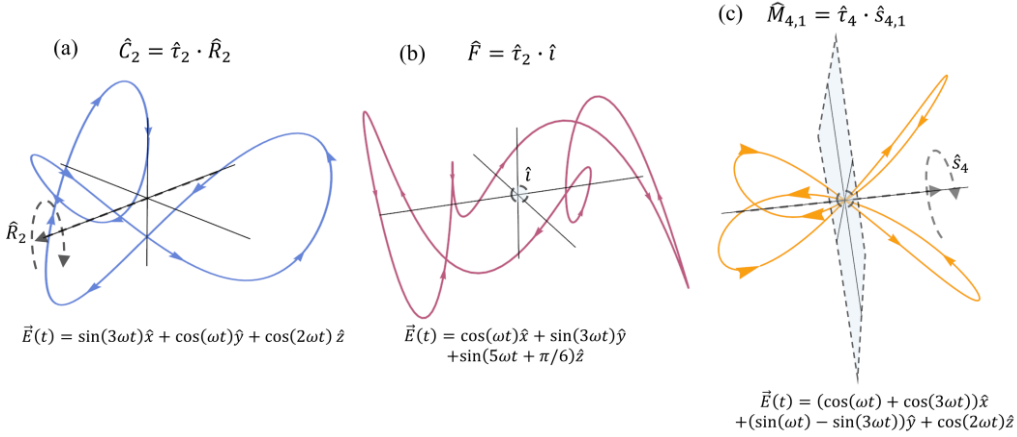

**Figure 5| (3+1)D DSs with examples for time-periodic fields possessing each symmetry. a** $\hat{C}_2$ symmetry, **b** $\hat{F}$ symmetry, and **c** $\hat{M}_{4,1}$ symmetry. The fields are represented on 3D Lissajou plots. The spatial parts of the operators are indicated by dashed arrows and planes, colored arrows along the plot indicate the direction of time.

**Dynamical symmetry selection rules**

We derive the physical constraints that arise in systems exhibiting DSs. We consider a general periodic time-dependent Hamiltonian: $H(t) = H(t + T)$. The Floquet Hamiltonian is defined as[47]:

$$\mathcal{H}_f = (H(t) - i\partial_t), \tag{20}$$

whose eigenstates are $T$-periodic Floquet modes ($|u_n(t)\rangle$), with corresponding quasi-energies ($\varepsilon_n$). Solutions to the time-dependent Schrödinger equation (TDSE) are comprised of Floquet states: $|\phi_n(t)\rangle = \exp(i\varepsilon_n t)|u_n(t)\rangle$, which we assume are nondegenerate. If $H(t)$ conforms to a DS group $G$, it commutes with all operators in $G$:

$$[\hat{X}, \mathcal{H}_f] = 0, \hat{X} \in G \tag{21}$$

In this case, the Floquet modes are simultaneous eigenmodes of the Floquet Hamiltonian, and of $\hat{X}$. Also, since $\hat{X}$ is unitary or anti-unitary, its eigenvalues are roots of unity: $\hat{X}|u_n(t)\rangle = e^{i\theta_n}|u_n(t)\rangle$, where $\theta_n$ is real. Thus, if the initial wave function populates a single Floquet state, the symmetry directly manifests in the time-dependent probability density $||\phi_n(t)\rangle|^2$ (or any other quantity that does not depend on the phase of the state). In this case, any measured observable $\vec{o}(t)$ also upholds the DS:

$$\vec{o}(t) = \langle u_n(t)|\hat{\vec{O}}|u_n(t)\rangle = \langle u_n(t)|\hat{X}^\dagger \cdot \hat{X} \cdot \hat{\vec{O}} \cdot \hat{X}^\dagger \cdot \hat{X}|u_n(t)\rangle = \hat{X} \cdot \vec{o}(t) \tag{22}$$

where $\hat{X}^\dagger$ is the inverse of $\hat{X}$. Any observable then complies to all the spatiotemporal symmetries in the group $G$, which imposes "selection rules" on its spectrum (Eq. (22) can be generalized to spatially dependent observables, e.g. Flux operators by transforming the spatially dependent part). In sections S.1 in the SI we show that Eq. (22) can be re-written as a set of eigenvalue problems in the Fourier domain. Also, in section S.2 we show that the conditions imposed on the temporal evolution can be expressed by the generating DSs in $G$ alone. If $G$ is non-abelian, $|u_n(t)\rangle$ can become degenerate, in which case Eq. (22) applies only to a set of commuting operators in $G$, of which the initial state is chosen as an eigenmode. Still, for an isotropic ensemble (that contains an equal population of all degenerate states), the observables are symmetric under all operations in $G$ even if it is non-abelian, since the sum of the projections on all degenerate states is always symmetric. $|u_n(t)\rangle$ may also become degenerate if $G$ contains time-reversal symmetry and the system has a non-integer total spin (through Kramers degeneracy[48]). This condition results from the anti-unitarity of time reversal, as opposed to spatial reflections which are unitary, hence this feature does not have an analogue in previously derived line-groups[30] and space-time groups[28]. In this case, Eq. (22) still holds for isotropic ensembles that populate an equal amount of the Kramers pairs (the degenerate time reversal partner states) due to the same conditions described above. Lastly, we note that in a homogenous chiral ensemble, any DSs that involve spatial reflections, inversions, or improper rotations will not lead to selection rules as a consequence of the medium breaking those symmetries.

**Selection rules for harmonic generation**

As an example, we apply the theory to HG, deriving the selection rules for important symmetries, and discussing several symmetry breaking mechanisms.

We begin by analyzing the symmetries of the HG Hamiltonian. Within the Born-Oppenheimer (BO) and dipole approximations, the microscopic Hamiltonian of a nonlinear medium interacting with a laser field is given in atomic units and in the length gauge by:

$$H_{HG}(t) = -\frac{1}{2}\sum_j \vec{\nabla}_j^2 + \frac{1}{2}\sum_{i\neq j}\frac{1}{|\vec{r}_i - \vec{r}_j|} + \sum_j V(\vec{r}_j) + \sum_j \vec{E}(t)\cdot\vec{r}_j \quad (23)$$

where $H_{HG}$ is the full time-dependent multi-electron Hamiltonian, $\vec{r}_j$ is the coordinate of the $j$'th electron, $\vec{\nabla}^2_j$ is the Laplacian operator with respect to $\vec{r}_j$, $V(\vec{r})$ is the potential energy term, $\vec{E}(t)$ is the electric driving laser field, and we neglect spin-orbit interactions. This Hamiltonian describes HG in most atomic, molecular and solid media (the theory can in principle be applied to any Hamiltonian, e.g. Hamiltonians that include spin-orbit terms, non-dipole light matter interaction terms, etc.) For a DS to result in selection rules, it should commute with the Hamiltonian, i.e., commute with all four terms in Eq. (23). Notably, the first three terms in $H_{HG}$ are time-independent, and therefore invariant under any temporal operation. Furthermore, the kinetic and electron-electron interaction terms are both invariant under all possible rotations and reflections. The spatial symmetries of $V(\vec{r})$ can be analyzed by standard group theory. It is then left to match the symmetries of $V(\vec{r})$ with the spatial part of the DSs. The remaining laser-matter interaction term is the only time-dependent term in $H_{HG}$, and the analysis of its symmetries is equivalent to analyzing the DSs of $\vec{E}(t)$. Overall, for a DS to commute with $H_{HG}$, the driver field should exhibit the DS, and the potential term should exhibit the spatial part of the DS.

In HG, the high harmonic spectrum is found by Fourier transforming the time-dependent polarization:

$$\vec{p}(t) = \langle\psi(t)|\hat{\vec{r}}|\psi(t)\rangle \quad (24)$$

where $\psi(t)$ is the solution to the TDSE. DSs that commute with the Hamiltonian impose constraints on $\vec{p}(t)$ (Eq. (22)), and lead to selection rules on the harmonic spectrum. To derive the selection rules, we analyze a general $\vec{p}(t)$ function that upholds the constraints in the Fourier-domain (section S.3 in the SI).

Table 1 presents selection rules for (2+1)D DSs in collinear HG. These include four novel selection rules: 1) a reflection time-reversal symmetry ($\hat{D}, \hat{H}$) that results in elliptically polarized harmonics, where the elliptical major or minor axis corresponds to the reflection axis. 2) When HG is driven by a laser field which is time-reversal ($\hat{T}$) invariant, the high harmonics are linearly polarized. Notably, such fields (e.g. see Fig. S.3 in the SI) can lead to rich two dimensional electron dynamics and re-collisions from multiple directions. Nonetheless, symmetry dictates that the different contributions interfere in a manner that leads to linearly polarized harmonics only. 3) A reflection translation symmetry ($\hat{Z}$) results in even harmonics that are linearly-polarized along the symmetry axis, and odd harmonics that are linearly polarized perpendicular to it, as recently demonstrated experimentally by HHG from cross-linear ω-2ω pumps[49,50]. Our results indicate that this geometry is a subset of a wider array of laser fields that uphold this symmetry (examples are shown in Fig. S.1 in the SI). 4) Elliptical symmetry ($\hat{e}_{n,m}$) results in a high harmonic spectrum where **all** harmonic orders have exactly the same ellipticity, which corresponds to the ellipticity parameter, $b$, of the Hamiltonian's underlying symmetry. The helicities alternate between harmonic orders, similar to the circular case ($\hat{C}_{n,m}$). Elliptically polarized high harmonics with fully tunable ellipticity may be useful for ultrafast spectroscopy and HG-based ellipsometry[51]. Notably, the fact that the ellipticity is determined by the DS makes it much more robust to perturbations than in other techniques[36].

**Table 1| (2+1)D DSs and their associated selection rules for collinear atomic/molecular HG.**

| Symmetry | Order | Harmonic generation selection rules |
|---|---|---|
| $\hat{D}, \hat{H}$ | 2 | Elliptically polarized harmonics with major/minor axis corresponding to the reflection axis. |
| $\hat{T}, \hat{Q}, \hat{G}$ | 2 | Linearly-polarized only harmonics. |
| $\hat{Z}$ | 2 | Linearly-polarized only harmonics, even harmonics are polarized along the reflection axis, and odd harmonics are polarized orthogonal to the reflection axis. |
| $\hat{C}_2$ | 2 | Odd-only harmonics, any polarization is possible. |
| $\hat{C}_{n,m}$ | n >2 | (±) circularly polarized ($nq \mp m$) harmonics, $q \in \mathbb{N}$, all other orders forbidden. |
| $\hat{e}_{n,m}$ | n >2 | (±) elliptically polarized ($nq \mp m$) harmonics, $q \in \mathbb{N}$, with an ellipticity $b$, all other orders forbidden. |

For the (3+1)D case that describes non-collinear geometries, DSs and their associated selection rules are presented in Table 2. We note two novel selection rules: 1) $n$-fold rotational DS ($\hat{e}_{n,m}$) leads to $nq$ harmonics that are all allowed (for integer $q$), suffice they are polarized only along the rotation axis. This can lead to photon mixing between all three polarization axes. 2) Improper rotational DSs ($\hat{P}_{n,m}$) lead to selection rules with symmetry-forbidden harmonics, as well as elliptically/circularly polarized harmonics. These are both numerically verified and discussed in section S.6 in the SI.

**Table 2| (3+1)D DSs and their associated selection rules for collinear or non-collinear atomic/molecular/solid HG.**

| Symmetry | Order | Harmonic generation selection rules |
|---|---|---|
| $\hat{D}, \hat{H}$ | 2 | The polarization ellipsoid has a major/minor axis normal to the reflection plane. |
| $\hat{Q}, \hat{G}$ | 2 | The rotation axis is a major/minor axis of the polarization ellipsoid. |
| $\hat{Z}$ | 2 | Odd harmonics are polarized linearly and orthogonally to the reflection plane, only even harmonics allowed polarized within the reflection plane |
| $\hat{C}_2$ | 2 | Odd-only harmonics in any polarization are allowed polarized in the plane orthogonal to rotation axis, any harmonic emission is allowed polarized parallel to the rotation axis. |
| $\hat{T}, \hat{J}, \hat{A}$ | 2 | Linearly polarized harmonics only. |
| $\hat{F}$ | 2 | Odd-only harmonics in any polarization. |
| $\hat{e}_{n,m}$ | n >2 | (±) elliptically polarized ($nq \mp m$) harmonics, $q \in \mathbb{N}$, with an ellipticity $b$ within the plane orthogonal to rotation axis. Linearly polarized $nq$ harmonics are also allowed, but polarized parallel to the rotation axis. |
| $\hat{P}_{2n,m}$ | 2n > 2 | (±) elliptically polarized ($2nq \mp m$) harmonics, $q \in \mathbb{N}$, with an ellipticity $b$ within the plane orthogonal to the improper rotation axis. $n(2q + 1)$ harmonics are also allowed, but polarized parallel to improper rotation axis. |
| $\hat{P}_{2n+1,m}$ | 2(2n + 1)>2 | (±) elliptically polarized $2q(2n + 1) \mp 2m$ harmonics, $q \in \mathbb{N}$, with an ellipticity $b$ within the plane orthogonal to the improper rotation axis. $(2n + 1)(2q + 1)$ harmonics are also allowed, but polarized parallel to improper rotation axis. |

The symmetry of the system (the Hamiltonian in Eq. (23)) is determined by the DSs of the pump field, and the symmetries of the potential $V(\vec{r})$. Often, HG is driven in atomic gas, and $V(\vec{r})$ is spherically symmetric; hence, selection rules are a consequence of only the DSs of the pump field (see section S.6 in SI). Still, there are various interesting cases when the medium is not spherically symmetric that are worth discussing. First, randomly distributed (non-oriented) molecular gas. In this case, the pump field interacts with all orientations of the molecule, and it is therefore the symmetry of the orientation averaged ensemble that affects the HG selection rules. The orientation averaged ensemble is either O(3) symmetric if it is achiral, or SO(3) symmetric if it is chiral. The O(3) group is spherically symmetric, so HG from non-oriented achiral media leads to the same selection rules as those observed from atomic spherically symmetric media[52]. Notably, the symmetries of the molecule are 'washed-out', and do not lead to selection rules in the HG spectrum (see section S.7 in SI). This picture is also valid for randomly distributed chiral media, except that SO(3) does not contain reflection symmetries, so reflectional based selection rules are broken (which can be used for chiral spectroscopy[53]). Second, a gas of aligned or oriented molecules[45,52,54–56]. Here, $V(\vec{r})$ depends on the relative orientation with respect to the laser field, and selection rules are only observed if both the medium and pump are co-aligned along the DS axis/plane (section S.7 in SI). Third, HG in solid media[57–60]. Some selection rules were previously derived for HG in solids, but these either dealt with rotational DSs[61], or considered rotation/inversion symmetries of the solid while assuming it interacts with a monochromatic pump[31,62]. Here we derived the general case that accounts for the DSs of the incident laser field as well as those of the solid, which requires the (3+1)D formulation given in Table 2 even for collinear cases (since the crystal structure can couple different spatial axes). Thus, the DSs and selection rules presented in Table 2 are new in solid HG (except for the $\hat{C}_n$ operator). Here the symmetries of the solid and the respective angle of incidence of the pump can determine the system's overall symmetry, which could be used for HG spectroscopy. For example, DS breaking spectroscopy can be used for orientation spectroscopy (see Fig S.9 in SI), as well as to probe spin-orbit and magnetic interactions (which break reflectional based DSs and time-reversal based DSs, respectively).

The analysis above assumes that the Hamiltonian is fully invariant under the DS transformations. In reality, HG Hamiltonians are not perfectly invariant. For instance, ionization and the driver pulse finite duration perturb time-reversal and translation symmetries. In fact, time-reversal symmetry can be significantly broken when the ionization rate per optical cycle is large, because field-induced tunneling is a non-reversible process (this effect is significant for harmonics above the ionization potential - see section S.6 in the SI). Still, selection rules in HG are observed routinely, both numerically and experimentally[26,36,52]. The actuality of the selection rule associated with the elliptical symmetry operation, $\hat{e}_{n,m}$, is more subtle, because the kinetic energy and potential terms are generally not invariant under the elliptical transformation. One approach to address this inconsistency is to use drivers that exhibit both spherical and elliptical DSs. A burst of linearly-polarized pump trains[63,64] and a synthetic piecewise driver pump can conform to this condition (S.6 in SI). Another approach is to generate harmonics in media with an elliptically symmetric kinetic term, e.g. by utilizing the concept of transformation optics[65]. It is also worth mentioning that propagation effects may influence the selection rules of macroscopic systems. For example, individual harmonic polarizations or intensities may change during propagation (but not forbidden harmonic selection rules). In fact, specially designed configurations may lead to generation of circularly polarized high harmonics, even though the emission from each atom is polarized linearly[66], or to quasi phase matching of only the even-order harmonics[67]. Notably, the presented theory can be extended to include propagation effects by applying the DS group theory approach also to Maxwell's equations and give rise to new selection rules in HG[68].

Lastly, we discuss the link between DSs and selection rules to conservation laws. Although Noether's theorem does not connect discrete DSs to conservation laws, all previous selection rules in HG were also derived from conservation laws: the appearance of only discrete harmonics in the spectrum (i.e., the selection rule due to time-periodicity $\hat{\tau}_1$) can be derived from energy conservation, the $\hat{C}_2$ selection rule can be derived due to parity conservation, and the $\hat{C}_n$ selection rule can be derived from conservation of spin angular momentum of the interacting photons[69,70]. This duality is believed to reflect an equivalence between the DS and photonic pictures in HG[69]. Interestingly, we find that several of the novel DSs ($\hat{e}_{n,m}, \hat{P}_{n,m}, \hat{Z}, \hat{T}, \hat{Q}, \hat{D},$ and $\hat{J}$) lead to selection rules that have no analogue conservation law derivation. What does this result mean? It probably shows that the DS perspective is more general than the photonic perspective. Can new conservation laws, associated with the above novel DSs, save the disparity between the two approaches?

**Conclusions and outlook**

In this paper we have implemented the concepts of group theory to map and characterize dynamical symmetries (DSs) exhibited by Floquet systems. We introduced novel symmetries that involve time reversal operations, and an elliptical symmetry that does not exist in molecular groups. We proved that if a given Hamiltonian commutes with a dynamical group then the generators of the group constrain the temporal evolution of Floquet states and the associated physical observables. We described these symmetry constraints as eigenvalue problems in the Fourier-domain, and derived their resulting selection rules.

For collinear atomic/molecular harmonic generation (HG), we discovered several selection rules: a reflection-time-reversal DS ($\widehat{D},\widehat{H}$) results in elliptically polarized harmonics with an elliptical major/minor axis parallel to the reflection axis. Any driver with time-reversal ($\widehat{T}$), or time-reversal followed by $\pi$ rotations ($\widehat{Q},\widehat{G}$) results in linearly-polarized harmonics. A spatial reflection followed by temporal translation ($\widehat{Z}$) results in linearly-polarized harmonics, where even harmonics are polarized along the symmetry axis, and odd harmonics are orthogonal to it[49,50]. An elliptical DS ($\hat{e}_{n,m}$) results in elliptically polarized eigenstates, and is useful for robust control of the ellipticity of all the high-order harmonics, collectively. These high harmonics would be useful for HG-based ellipsometry[51]. Furthermore, we explored non-collinear and solid HG and found that these include new DSs that do not exist in the (2+1)D case: an inversion-time-translation DS that leads to odd only harmonics ($\widehat{F}$), an inversion-time-reversal DS that leads to linearly polarized harmonics ($\widehat{J},\widehat{A}$), and improper-rotational DSs that lead to circularly/elliptically polarized high harmonics within the symmetry plane, and linearly polarized harmonics orthogonal to it ($\widehat{M}/\widehat{P}$). Interestingly, in contrast to current selection rules that can be derived by DSs or equivalently from conservation laws, some of the new selection rules cannot be derived by the known conservation laws.

A very exciting aspect of the HG group theory is that it forms a starting point for utilizing and investigating broken symmetries. In this respect, even if a DS leads to a 'boring' selection rule (e.g. time-reversal DS, $\widehat{T}$, that 'just' leads to a linearly polarized spectrum), it still has major implications for ultrafast spectroscopy of various systems. In this sense, there are no boring selection rules. For example, the HG group theory can be used to characterize the symmetries and determine the orientation of any molecular or solid medium[52] (see sections S.7, S.8 in SI), time-translation/time-reversal DSs ($\hat{t}_n, \widehat{T}$) can be used to find and characterize atomic/molecular resonances and probe ionization dynamics (see section S.6 in SI), time-reversal DSs ($\widehat{T}, \widehat{Q}, \widehat{D}, \widehat{J}$) can be used to probe magnetic interactions, reflection DSs ($\widehat{Z}, \widehat{D}$) can shed light on spin-orbit interactions in atomic, molecular, and solid systems, and as will be reported soon[53], reflection/inversion DSs ($\widehat{Z}, \widehat{D}, \widehat{J}, \widehat{F}$) can be used to probe dynamical chiral processes[71], and more.

We applied here the derived Floquet group theory for analyzing HG at the microscopic level. Its application for exploring wave propagation effects will follow soon[68]. Our work also paves the way for several interesting directions beyond nonlinear optics. The introduced elliptical DS can be implemented in other systems, both static (e.g. metamaterials[6,7] and transformation optics[65]) and Floquet, to yield elliptical eigenstates. Extending the theory to lattice systems by including translational operators gives rise to novel 'dynamical space groups', which may lead to group theory based classification of Floquet topological insulators[10,13], time-crystals[72], and "shaken" optical lattices[16,17]. Extensions to non-Hermitian systems, e.g. *PT* symmetric waveguides[73], and quasi-periodic Floquet systems[74] should also be possible and exciting, leading to new DSs and selection rules. Overall, we expect that group theory analysis of DSs will lead to extended understanding and novel discoveries in various Floquet systems.

## Acknowledgements

This work was supported by the Israel Science Foundation (grant no. 1225/14 and 1839/13), the Israeli Center of Research Excellence 'Circle of Light' supported by the I-CORE Program of the Planning and Budgeting Committee and the Israel Science Foundation (grant no. 1802/12), the joint UGC-ISF Research Grant Program (grant no. 1903/14), and the Wolfson foundation. O.N. gratefully acknowledges the support of the Adams Fellowship Program of the Israel Academy of Sciences and Humanities.

## References

1.  Weyl, H. *The theory of groups and quantum mechanics*. (Courier Corporation, 1950).
2.  Weyl, H. *Symmetry*. (Princeton University, Princeton, NJ, 1989).
3.  Harris, D. C. & Bertolucci, M. D. *Symmetry and spectroscopy: an introduction to vibrational and electronic spectroscopy*. (Courier Corporation, 1978).

4. Bishop, D. M. *Group theory and chemistry*. (Courier Corporation, 2012).
5. Simon, H. J. & Bloembergen, N. Second-Harmonic Light Generation in Crystals with Natural Optical Activity. *Phys. Rev.* **171,** 1104–1114 (1968).
6. Konishi, K. *et al.* Polarization-controlled circular second-harmonic generation from metal hole arrays with threefold rotational symmetry. *Phys. Rev. Lett.* **112,** 1–5 (2014).
7. Chen, S. *et al.* Symmetry selective third harmonic generation from plasmonic metacrystals. *Phys. Rev. Lett.* **113,** 13 (2014).
8. McPherson, A. *et al.* Studies of multiphoton production of vacuum-ultraviolet radiation in the rare gases. *J. Opt. Soc. Am. B* **4,** 595–601 (1987).
9. Krausz, F. & Ivanov, M. Attosecond physics. *Rev. Mod. Phys.* **81,** 163–234 (2009).
10. Lindner, N. H., Refael, G. & Galitski, V. Floquet Topological Insulator in Semiconductor Quantum Wells. *Nat. Phys.* **7,** 490–495 (2010).
11. Kitagawa, T., Oka, T., Brataas, A., Fu, L. & Demler, E. Transport properties of nonequilibrium systems under the application of light: Photoinduced quantum Hall insulators without Landau levels. *Phys. Rev. B* **84,** (2011).
12. Lindner, N. H., Bergman, D. L., Refael, G. & Galitski, V. Topological Floquet spectrum in three dimensions via a two-photon resonance. *Phys. Rev. B* **87,** 1–11 (2013).
13. Rechtsman, M. C. *et al.* Photonic Floquet topological insulators. *Nature* **496,** 196–200 (2013).
14. Wang, R., Wang, B., Shen, R., Sheng, L. & Xing, D. Y. Floquet Weyl semimetal induced by off-resonant light. *EPL (Europhysics Lett.* **105,** 17004 (2014).
15. Wang, H., Zhou, L. & Chong, Y. D. Floquet Weyl phases in a three-dimensional network model. *Phys. Rev. B* **93,** 144114 (2016).
16. Eckardt, A., Weiss, C. & Holthaus, M. Superfluid-insulator transition in a periodically driven optical lattice. *Phys. Rev. Lett.* **95,** 260404 (2005).
17. Grushin, A. G., Gómez-León, Á. & Neupert, T. Floquet fractional Chern insulators. *Phys. Rev. Lett.* **112,** 156801 (2014).
18. Zhang, P. & Zhao, X.-G. Quantum dynamics of a driven double quantum dot. *Phys. Lett. A* **271,** 419–428 (2000).
19. Platero, G., Go, A., Gómez-León, A. & Platero, G. Floquet-Bloch theory and topology in periodically driven lattices. *Phys. Rev. Lett.* **110,** 200403 (2013).
20. Delplace, P., Gómez-León, Á. & Platero, G. Merging of Dirac points and Floquet topological transitions in ac-driven graphene. *Phys. Rev. B* **88,** 245422 (2013).
21. Titum, P., Lindner, N. H., Rechtsman, M. C. & Refael, G. Disorder-Induced Floquet Topological Insulators. *Phys. Rev. Lett.* **114,** 56801 (2015).
22. Gemelke, N., Sarajlic, E., Bidel, Y., Hong, S. & Chu, S. Parametric amplification of matter waves in periodically translated optical lattices. *Phys. Rev. Lett.* **95,** 2–5 (2005).
23. Lignier, H. *et al.* Dynamical control of matter-wave tunneling in periodic potentials. *Phys. Rev. Lett.* **99,** 1–4 (2007).
24. Weitz, M. *et al.* Tuning the mobility of a driven Bose-Einstein condensate via diabatic Floquet bands. *Phys. Rev. Lett.* **110,** 135302 (2013).
25. Ben-Tal, N., Moiseyev, N. & Beswick, A. The effect of Hamiltonian symmetry on generation of odd and even harmonics. *J. Phys. B At. Mol. Opt. Phys.* **26,** 3017 (1993).
26. Alon, O. E., Averbukh, V. & Moiseyev, N. Selection Rules for the High Harmonic Generation Spectra. *Phys. Rev. Lett.* **80,** 3743–3746 (1998).
27. Averbukh, V., Alon, O. & Moiseyev, N. Crossed-beam experiment: High-order harmonic generation and dynamical symmetry. *Phys. Rev. A* **60,** 2585–2586 (1999).
28. Janssen, T., Janner, A. & Ascher, E. Crystallographic groups in space and time. *Physica* **41,** 541–565 (1969).
29. Vujicic, M., Bozovic, I. B. & F, H. Construction of the symmetry groups of polymer molecules. **10,**

1271–1279 (1977).
30. Damnjanovic, M. & Milošević, I. *Line Groups in Physics: Theory and Applications to Nanotubes and Polymers (Lecture Notes in Physics vol 801)*. (Berlin: Springer, 2010).
31. Yariv, A. & Yeh, P. *Photonics: optical electronics in modern communications*. **6,** (Oxford University Press New York, 2007).
32. Eichmann, H. *et al.* Polarization-dependent high-order two-color mixing. *Phys. Rev. A* **51,** R3414(R) (1995).
33. Milošević, D. B., Becker, W. & Kopold, R. Generation of circularly polarized high-order harmonics by two-color coplanar field mixing. *Phys. Rev. A* **61,** 1–15 (2000).
34. Milošević, D. B. & Becker, W. Attosecond pulse trains with unusual nonlinear polarization. *Phys. Rev. A* **62,** 11403 (2000).
35. Averbukh, V., Alon, O. & Moiseyev, N. Stability and instability of dipole selection rules for atomic high-order-harmonic-generation spectra in two-beam setups. *Phys. Rev. A* **65,** 15–17 (2002).
36. Fleischer, A., Kfir, O., Diskin, T., Sidorenko, P. & Cohen, O. Spin angular momentum and tunable polarization in high-harmonic generation. *Nat. Photonics* **8,** 543–549 (2014).
37. Kfir, O. *et al.* Generation of bright phase-matched circularly-polarized extreme ultraviolet high harmonics. *Nat. Photonics* **9,** 99–105 (2014).
38. Milošević, D. B. Circularly polarized high harmonics generated by a bicircular field from inert atomic gases in the p state: A tool for exploring chirality-sensitive processes. *Phys. Rev. A* **92,** 43827 (2015).
39. Uzer, F. M. and A. D. B. and T., Mauger, F., Bandrauk, A. D. & Uzer, T. Circularly polarized molecular high harmonic generation using a bicircular laser. *J. Phys. B At. Mol. Opt. Phys.* **49,** 10LT01 (2016).
40. Bandrauk, A. D., Mauger, F. & Yuan, K.-J. Circularly polarized harmonic generation by intense bicircular laser pulses: electron recollision dynamics and frequency dependent helicity. *J. Phys. B At. Mol. Opt. Phys.* **49,** 23LT01 (2016).
41. Ceccherini, F., Bauer, D. & Cornolti, F. Dynamical symmetries and harmonic generation. *J. Phys. B At. Mol. Opt. Phys.* **34,** 5017–5029 (2001).
42. Ceccherini, F. & Bauer, D. Harmonic generation in ring-shaped molecules. *Phys. Rev. A. At. Mol. Opt. Phys.* **64,** 1–7 (2001).
43. Averbukh, V., Alon, O. E. & Moiseyev, N. High-order harmonic generation by molecules of discrete rotational symmetry interacting with circularly polarized laser field. *Phys. Rev. a* **6403,** art. no.-033411 (2001).
44. Ofir E. Alon. Dynamical symmetries of time-periodic Hamiltonians. *Phys. Rev. A* **66,** 13414 (2002).
45. Liu, X. *et al.* Selection rules of high-order-harmonic generation: Symmetries of molecules and laser fields. *Phys. Rev. A* **94,** 1–9 (2016).
46. Fan, T. *et al.* Bright circularly polarized soft X-ray high harmonics for X-ray magnetic circular dichroism. *Proc. Natl. Acad. Sci. U. S. A.* **112,** 14206–11 (2015).
47. Holthaus, M. Floquet engineering with quasienergy bands of periodically driven optical lattices. *J. Phys. B At. Mol. Opt. Phys.* **49,** 13001 (2016).
48. Kramers, H. A. General theory of paramagnetic rotation in crystals. in *Proc. Acad. Sci. Amsterdam* **33,** 959 (1930).
49. Dudovich, N. *et al.* Measuring and controlling the birth of attosecond XUV pulses. *Nat. Phys.* **2,** 781–786 (2006).
50. Shafir, D., Mairesse, Y., Villeneuve, D. M., Corkum, P. B. & Dudovich, N. Atomic wavefunctions probed through strong-field light–matter interaction. *Nat. Phys.* **5,** 412–416 (2009).
51. Brimhall, N. *et al.* Characterization of optical constants for uranium from 10 to 47 nm. *Appl. Opt.* **49,** 1581–1585 (2010).
52. Baykusheva, D., Ahsan, M. S., Lin, N. & Wörner, H. J. Bicircular High-Harmonic Spectroscopy Reveals Dynamical Symmetries of Atoms and Molecules. *Phys. Rev. Lett.* **116,** 123001 (2016).
53. Neufeld, O. & Cohen, O. Highly selective chiral discrimination in high harmonic generation by

dynamical symmetry breaking spectroscopy. *Submitted* (arXiv:1807.02630)

54. Bandrauk, A. D. & Lu, H. Controlling harmonic generation in molecules with intense laser and static magnetic fields: Orientation effects. *Phys. Rev. A* **68,** 43408 (2003).
55. Žďánská, P., Averbukh, V. & Moiseyev, N. High harmonic generation spectra of aligned benzene in circular polarized laser field. *J. Chem. Phys.* **118,** 8726 (2003).
56. Hasović, E., Odžak, S., Becker, W. & Milošević, D. B. High-order harmonic generation in non-planar molecules driven by a bicircular field. **8976,** (2016).
57. Ghimire, S. *et al.* Observation of high-order harmonic generation in a bulk crystal. *Nat. Phys.* **7,** 138 (2010).
58. Higuchi, T., Stockman, M. I. & Hommelhoff, P. Strong-Field Perspective on High-Harmonic Radiation from Bulk Solids. *Phys. Rev. Lett.* **113,** 213901 (2014).
59. McDonald, C. R., Vampa, G., Corkum, P. B. & Brabec, T. Interband Bloch oscillation mechanism for high-harmonic generation in semiconductor crystals. *Phys. Rev. A* **92,** 33845 (2015).
60. Saito, N. *et al.* Observation of selection rules for circularly polarized fields in high-harmonic generation from a crystalline solid. *Optica* **4,** 1333–1336 (2017).
61. Alon, O., Averbukh, V. & Moiseyev, N. High harmonic generation of soft X-rays by carbon nanotubes. *Phys. Rev. Lett.* **85,** 5218–21 (2000).
62. Tang, C. L. & Rabin, H. Selection Rules for Circularly Polarized Waves in Nonlinear Optics. *Phys. Rev. B* **3,** 4025–4034 (1971).
63. Neufeld, O., Bordo, E., Fleischer, A. & Cohen, O. High harmonic generation with fully tunable polarization by train of linearly-polarized pulses. *New J. Phys.* **19,** (2017).
64. Neufeld, O., Bordo, E., Fleischer, A. & Cohen, O. High Harmonics with Controllable Polarization by a Burst of Linearly-Polarized Driver Pulses. *Photonics* **4,** 31 (2017).
65. Pendry, J. B., Schurig, D. & Smith, D. R. Controlling Electromagnetic Fields. *Science.* **312,** 1780–1782 (2006).
66. Liu, L. Z., O'Keeffe, K. & Hooker, S. M. Optical rotation quasi-phase-matching for circularly polarized high harmonic generation. *Opt. Lett.* **37,** 2415–2417 (2012).
67. Diskin, T. & Cohen, O. Quasi-phase-matching of only even-order high harmonics. *Opt. Express* **22,** 7145–53 (2014).
68. Lerner, G., Hareli, L., Neufeld, O., Fleischer, A., Bahabad. A. & Cohen, O. Quasi-phase matching selection rules by Flouqet group theory. *In preperation*
69. Milošević, D. B. High-order harmonic generation by a bichromatic elliptically polarized field: conservation of angular momentum. *J. Phys. B At. Mol. Opt. Phys.* **48,** 171001 (2015).
70. Kfir, O. *et al.* Helicity-selective phase-matching and quasi-phase matching of circularly polarized high-order harmonics: towards chiral attosecond pulses. *J. Phys. B At. Mol. Opt. Phys.* **49,** 123501 (2016).
71. Cireasa, R. *et al.* Probing molecular chirality on a sub-femtosecond timescale. *Nat. Phys.* **11,** 654 (2015).
72. Else, D. V, Bauer, B. & Nayak, C. Floquet Time Crystals. *Phys. Rev. Lett.* **117,** 90402 (2016).
73. Klaiman, S., Günther, U. & Moiseyev, N. Visualization of Branch Points in PT-Symmetric Waveguides. *Phys. Rev. Lett.* **101,** 80402 (2008).
74. Gommers, R., Denisov, S. & Renzoni, F. Quasiperiodically Driven Ratchets for Cold Atoms. *Phys. Rev. Lett.* **96,** 240604 (2006).

# Supplementary information
# Symmetries and selection rules in Floquet systems: application to harmonic generation in nonlinear optics

*Ofer Neufeld[1,2], Daniel Podolsky[2] and Oren Cohen[1,2]*
[1]Solid state Institute, Technion - Israel Institute of Technology, Haifa 32000, Israel.
[2]Physics Department, Technion - Israel Institute of Technology, Haifa 32000, Israel.

The supplementary information file contains the following sections: in section S.1 we show that the selection rules arising from dynamical symmetries (DSs) can be treated as eigenvalue problems in the frequency-domain. In section S.2 we prove that the generating operators of a dynamical group are sufficient to describe the selection rules in Floquet systems. Section S.3 derives the selection rules for the harmonic generation (HG) spectra. Section S.4 provides a list of selected dynamical groups as developed in the main text. Section S.5 discusses similarities between Floquet groups and other groups. Lastly, sections S.6, S.7, and S.8, present numerical time dependent Schrödinger equation (TDSE) HG calculations for novel DSs (both in (2+1)D and (3+1)D), from atomic gas, oriented and non-oriented molecular gas, and solids, demonstrating that the analytically derived selection rules are upheld, and exploring DS breaking spectroscopy.

## S.1 DS constraints as eigenvalue problems

We show here how each DS constraint on a physical observable (derived in Eq. (22) in the main text) can be written as a set of eigenvalue problems in the Fourier domain. In order to understand how this takes place, we first demonstrate it for a general observable $\vec{o}(t)$ in a (2+1)D system with $\hat{C}_2 = \hat{\tau}_2 \cdot \hat{R}_2$ DS. Since $\vec{o}(t)$ is $T$-periodic ($T$-periodicity is a symmetry of all Floquet systems), it can be expanded to a Fourier series:

$$\vec{o}(t) = \sum_n \vec{F}_n \exp\left(in\frac{2\pi t}{T}\right) \quad (1)$$

where the coefficients $\vec{F}_n$ are complex numbers. Thus, the Fourier transform of $\vec{o}(t)$ is discrete in the frequency domain, with Dirac delta functions around integer multiples of the frequency $\omega_0 = 2\pi/T$. Furthermore, if $\vec{o}(t)$ is a real quantity, then the coefficients real parts are even, and their imaginary parts are odd:

$$\vec{F}_n = \vec{a}_n + i\vec{b}_n; \vec{a}_n = \vec{a}_{-n}, \vec{b}_n = -\vec{b}_{-n} \quad (2)$$

where $\vec{a}_n$ and $\vec{b}_n$ are real. Applying the symmetry constraint on Eq. (1), i.e. requiring that the observable is invariant to $\hat{C}_2$, yields:

$$\sum_n \hat{C}_2 \cdot \vec{F}_n \exp\left(in\frac{2\pi t}{T}\right) = \sum_n \vec{F}_n \exp\left(in\frac{2\pi t}{T}\right) \quad (3)$$

The coefficients of identical exponents in Eq. (3) must be equal. This reduces Eq. (3) to separate eigenvalue problems for each frequency component $n$:

$$\hat{R}_2 \cdot \vec{F}_n = e^{in\pi}\vec{F}_n \quad (4)$$

The rotational operation $\hat{R}_2$ has two identical eigenvalues of value -1, with any eigenvector. The constraints in Eq. (4) are upheld if the temporal part of the symmetry (that translates by half a period) leads to an eigenvalue that corresponds to those of the matrix $\hat{R}_2$, which only happens for odd values of $n$. That is, the equation is solved by any eigenvector for odd values of $n$, and has only a trivial solution for even values of $n$. All DSs can be treated in a similar manner, while only the eigenvalues and eigenvectors that uphold the equations change for different DSs. These constraints restrict the spectrum of the physical observable, and therefore its temporal evolution as well. A solution is always guaranteed: either a non-trivial one, or a trivial one ($\vec{F}_n = 0$). For HG, a trivial solution results in "forbidden harmonics", that is, selection rules. More generally, selection rules can be expressed as any relation between the components of $\vec{F}_n$, determined by the eigenvectors themselves. A similar derivation for a $\hat{C}_3$ symmetry which is present in counter-rotating bi-circular $\omega$-$2\omega$ laser fields, was presented in in ref. 1.

## S.2 Selection rules in groups with multiple members

In this section we prove that the generators of a symmetry group are sufficient to describe the selection rules and other requirements in Floquet systems. This is intuitive, since the generators provide all the information about the group (other members in the group result in restrictions that are upheld by closure and associativity in the group). First, we consider a group $G$ with a single generator, $\hat{X}$. The generator can be broken to spatial and temporal components:

$$\hat{X} = \hat{X}_{temp} \cdot \hat{X}_{space} \tag{5}$$

The resulting eigenvalue problems that represent the symmetry constraints due to $\hat{X}$ can be written as:

$$\sum_n \hat{X}_{space} \cdot \vec{F}_n \cdot \hat{X}_{temp} \cdot \exp\left(in\frac{2\pi t}{T}\right) = \sum_n \vec{F}_n \cdot \exp\left(in\frac{2\pi t}{T}\right) \tag{6}$$

The temporal part determines the eigenvalues, and the spatial part determines the operator to be diagonalized. A non-trivial solution exists only if the spatial and the temporal parts exactly match. Assuming a solution $\vec{F}_n$ that solves Eq. (6), symmetry operations that are powers of $\hat{X}$ lead to eigenvalue problems with the same solutions. The $m$'th power leads to:

$$\sum_n (\hat{X}_{space})^m \cdot \vec{F}_n \cdot (\hat{X}_{temp})^m \cdot \exp\left(in\frac{2\pi t}{T}\right) = \sum_n \vec{F}_n \cdot \exp\left(in\frac{2\pi t}{T}\right) \tag{7}$$

where we used the commutativity of the spatial and temporal operators. Due to associativity in the group, the equation above is solved by the same $\vec{F}_n$, with the eigenvalues of $\hat{X}$ to the power of $m$. Thus, symmetry constraints resulting from powers of $\hat{X}$ coincide with those imposed by $\hat{X}$.

Next, we consider two generators, $\hat{X}$ and $\hat{Y}$ and their resulting symmetry constraints. Again, we assume a solution $\vec{F}_n$ that simultaneously solves the eigenvalue problems of $\hat{X}$ and $\hat{Y}$ combined. In this case, the same solution also solves the eigenvalue problem defined by the symmetry operation that is $\hat{X} \cdot \hat{Y}$:

$$\sum_n (\hat{X}_{space} \cdot \hat{Y}_{space}) \cdot \vec{F}_n \cdot (\hat{X}_{temp} \cdot \hat{Y}_{temp}) \cdot \exp\left(in\frac{2\pi t}{T}\right) = \sum_n \vec{F}_n \cdot \exp\left(in\frac{2\pi t}{T}\right) \tag{8}$$

where we again used the commutativity of the spatial and temporal operators. Eq. (8) can be thought of as a series multiplication, or as a single eigenvalue problem whose eigenvalues are multiplications of those of $\hat{X}$ and $\hat{Y}$. Clearly, if all symmetry constraints due to powers of $\hat{X}$ and $\hat{Y}$ and their products are upheld, then the same solution is valid for all symmetry constraints in the group, since the group is comprised of powers and products of its generators. At this point the theorem is trivially extended to any number of generators, and results in the following: given a dynamical group $G$, one must only examine the selection rules arising from the generating symmetries.

## S.3 HG selection rules

We develop the HG selection rules for underlying DSs using the eigenvalue problem approach described in section S.1 for the time-dependent polarization (Eq. (24) in the main text), which is even under time-reversal (odd observables, e.g. currents, have slightly different selection rules). We begin with the (2+1)D case that describes DSs in collinear atomic/molecular HG, and then move on to the (3+1)D case that describes non-collinear and solid HG.

### S.3.1 HG selection rules in (2+1)D

**Time reversal symmetry - $\hat{T}$**

$\hat{T}$ symmetry results in the following constraints:

$$\vec{F}_n = \vec{F}_{-n} \tag{9}$$

A general solution demands the $x$ and $y$ components of $\vec{F}_n$ are in phase, meaning any driver invariant under time-reversal results in linearly-polarized only harmonics.

**Reflection-time reversal symmetry - $\hat{D}_x$**

$\widehat{D}_x$ symmetry results in the following constraints:

$$\hat{\sigma}_x \cdot \vec{F}_n = \vec{F}_{-n} \tag{10}$$

The selection rules dictate that the $x$ and $y$ components have a relative phase of $\pi/2$. This means any elliptical emission is allowed, as long as the polarization ellipse has a major/minor axis parallel to the reflection axis.

**Reflection-time translation symmetry - $\widehat{Z}_x$**

$\widehat{Z}_x$ symmetry results in the following constraints:

$$\hat{\sigma}_x \cdot \vec{F}_n = e^{in\pi}\vec{F}_n \tag{11}$$

Here we must separate to even and odd $n$ cases. The general solution dictates selection rules where even harmonics are linearly-polarized along the symmetry axis, and odd harmonics are linearly-polarized orthogonally to the symmetry axis. This selection rule was demonstrated experimentally for a cross-linear ω-2ω pumps driving HHG in refs. 2,3.

**Reflection-time translation followed by time reversal symmetry - $\widehat{H}_x$**

$\widehat{H}_x$ symmetry results in the following constraints:

$$\hat{\sigma}_x \cdot \vec{F}_n = e^{in\pi}\vec{F}_{-n} \tag{12}$$

The resulting selection rules dictate that the $x$ and $y$ components have a relative phase of $\pi/2$. That is, any elliptical emission is allowed, as long as the polarization ellipse has a major/minor axis parallel to the reflection axis. The identical selection rules to the operator $\widehat{D}_x$ are not coincidental, and indicate that it does not matter which point in time upholds $\widehat{D}_x$ symmetry (the additional time-translation operator shifts the $\hat{S}_x$ symmetry by $T/2$).

**Rotation by 180⁰-time translation symmetry - $\widehat{C}_2$**

$\widehat{C}_2$ symmetry was previously found responsible for odd-only harmonic emission from one-color laser fields[4], as we also show here. $\widehat{C}_2$ symmetry results in the following constraints:

$$\hat{R}_2 \cdot \vec{F}_n = e^{in\pi}\vec{F}_n \tag{13}$$

The spatial rotation operator by 180° degrees has two identical eigenvalues of value -1. A non-trivial solution then only exists for odd values of $n$. This results in odd-only harmonics, compatible with previous derivations, with no particular limitation on the polarization. For instance, slightly elliptical fields uphold this symmetry and produce elliptically polarized odd harmonics[5,6].

**Rotation by 180⁰-time reversal symmetry - $\widehat{Q}$**

$\widehat{Q}$ symmetry results in the following constraints:

$$\hat{R}_2 \cdot \vec{F}_n = \vec{F}_{-n} \tag{14}$$

The resulting selection rules dictate that the $x$ and $y$ components of $\vec{F}_n$ are in phase. Thus, any system upholding $\widehat{Q}$ symmetry generates only linearly-polarized harmonics.

**Rotation by 180⁰-time translation followed by time reversal symmetry - $\widehat{G}$**

$\widehat{G}$ symmetry results in the following constraints:

$$\hat{R}_2 \cdot \vec{F}_n = e^{in\pi}\vec{F}_{-n} \tag{15}$$

For which the resulting selection rules are identical to those imposed by $\widehat{Q}$ symmetry, with similar arguments.

**Rotation-time translation symmetry - $\widehat{C}_N$**

$\widehat{C}_N$ symmetry results in the following constraints:

$$\hat{R}_N \cdot \vec{F}_n = e^{in\frac{2\pi}{N}}\vec{F}_n \tag{16}$$

The rotational operator has two eigenvalues: $e^{\pm i2\pi/N}$, with the eigenvectors: $\begin{pmatrix}1\\\pm i\end{pmatrix}$. The selection rules dictate only circularly polarized $n = Nq \pm 1$ harmonics are emitted (for integer $q$), with alternating helicities. This derivation appeared in ref. 7.

**Generalized rotation-time translation symmetry - $\widehat{C}_{N,m}$**

This symmetry is exhibited in HHG driven by generalized two-color bi-circular EM fields[8,9]. $\widehat{C}_{N,m}$ symmetry results in the following constraints:

$$\hat{R}_{N,m} \cdot \vec{F}_n = e^{in\frac{2\pi}{N}} \vec{F}_n \qquad (17)$$

The resulting selection rules dictate only circularly polarized $n = Nq \pm m$ harmonics are emitted (for integer $q$), with alternating helicities, consistent with previous derivations from conservation laws[8,9].

**Generalized elliptical rotation-time translation symmetry - $\hat{e}_{N,m}$**

$\hat{e}_{N,m}$ symmetry results in the following constraints:

$$\hat{L}_b \cdot \hat{R}_{N,m} \cdot \hat{L}_{1/b} \cdot \vec{F}_n = e^{in\frac{2\pi}{N}} \vec{F}_n \qquad (18)$$

The generalized rotation-scaling operator has the same eigenvalues as the standard rotational operator, but different eigenvectors that depend on the ellipticity: $\begin{pmatrix} 1 \\ \pm bi \end{pmatrix}$. The resulting selection rules dictate that the emitted harmonics are elliptically polarized along the elliptical axis with alternating helicity, with the same underlying ellipticity $b$, and only $n = Nq \pm m$ harmonics are emitted (for integer $q$).

### S.3.2 HG Selection rules in (3+1)D

The (3+1)D formulation is identical to the (2+1)D case, except that rotations must now revolve around a specific axis in 3D space (which we always assume is the z-axis), and reflections are no longer reflection axes, but reflection planes (which we always assume to be the xy plane). Additionally, $\vec{F}_n$ is now a 3D object.

**Time reversal symmetry - $\hat{T}$**

$\hat{T}$ symmetry results in the following constraints:

$$\vec{F}_n = \vec{F}_{-n} \qquad (19)$$

A general solution demands the $x$, $y$, and $z$ components of $\vec{F}_n$ are in phase, meaning any driver invariant under time-reversal results in linearly-polarized only harmonics, same as the (2+1)D case.

**Reflection-time reversal symmetry - $\hat{D}_{xy}$**

$\hat{D}_{xy}$ symmetry results in the following constraints:

$$\hat{\sigma}_{xy} \cdot \vec{F}_n = \vec{F}_{-n} \qquad (20)$$

The selection rules dictate that the $x$ and $y$ components are in phase (i.e. a linear polarization within the xy plane), and are phase shifted by $\pi/2$ for the z-axis induced polarization. This means the 3D polarization ellipsoid has a 2D projection for each harmonic with a major/minor axis along the z-axis.

**Reflection-time translation symmetry - $\hat{Z}_{xy}$**

$\hat{Z}_{xy}$ symmetry results in the following constraints:

$$\hat{\sigma}_{xy} \cdot \vec{F}_n = e^{in\pi} \vec{F}_n \qquad (21)$$

Here we must separate to even and odd $n$ cases. The general solution dictates selection rules where even harmonics are in any elliptical polarization within the symmetry plane, and odd harmonics must be linearly-polarized orthogonally to the symmetry plane.

**Reflection-time translation followed by time reversal symmetry - $\hat{H}_{xy}$**

$\hat{H}_{xy}$ symmetry results in the following constraints:

$$\hat{\sigma}_{xy} \cdot \vec{F}_n = e^{in\pi} \vec{F}_{-n} \qquad (22)$$

The resulting selection rules are identical to those of the operator $\hat{D}_x$.

**Rotation by $180^0$-time translation symmetry - $\hat{C}_2$**

$\hat{C}_2$ symmetry results in the following constraints:

$$\hat{R}^z{}_2 \cdot \vec{F}_n = e^{in\pi} \vec{F}_n \qquad (23)$$

The selection rules lead to odd-only harmonics that can be polarized within the plane orthogonal to the rotation (xy plane), and only even harmonics polarized parallel to the symmetry axis (z-axis).

**Rotation by $180^0$-time reversal symmetry - $\hat{Q}$**

$\hat{Q}$ symmetry results in the following constraints:

$$\hat{R}_2 \cdot \vec{F}_n = \vec{F}_{-n} \qquad (24)$$

The resulting selection rules dictate that the $x$ and $y$ polarization components are in phase (i.e. linear polarization within the xy plane), and phase shifted by $\pi/2$ from the z symmetry axis. Thus, the 3D polarization ellipsoid has a 2D projection for each harmonic with a major/minor axis along the z-axis.

**Rotation by $180^0$-time translation followed by time reversal symmetry - $\hat{G}$**

$\hat{G}$ symmetry results in the following constraints:

$$\hat{R}_2 \cdot \vec{F}_n = e^{in\pi} \vec{F}_{-n} \qquad (25)$$

For which the resulting selection rules are identical to those imposed by $\hat{Q}$ symmetry.

**Inversion-time translation symmetry - $\hat{F}$**

$\hat{F}$ symmetry results in the following constraints:

$$\hat{\imath} \cdot \vec{F}_n = e^{in\pi} \vec{F}_n \qquad (26)$$

The selection rules lead to odd-only harmonics with any generalized polarization. We note this symmetry is well-known from perturbative nonlinear optics, and often attributed to centro-symmetric media which cancel out even order nonlinear coefficients[10]. In our case, $\hat{F}$ symmetry is the generalization of this selection rule, which also depends on the DS of the incident pump field.

**Inversion-time reversal symmetry - $\hat{J}$**

$\hat{J}$ symmetry results in the following constraints:

$$\hat{\imath} \cdot \vec{F}_n = \vec{F}_{-n} \qquad (27)$$

The resulting selection rules dictate that the $x$, $y$ and $z$ axes induced polarization components are in phase (i.e. any linearly polarized high harmonic is allowed).

**Inversion-time translation followed by time reversal symmetry - $\hat{A}$**

$\hat{A}$ symmetry results in the following constraints:

$$\hat{\imath} \cdot \vec{F}_n = e^{in\pi} \vec{F}_{-n} \qquad (28)$$

For which the resulting selection rules are identical to those imposed by $\hat{J}$ symmetry.

**Generalized elliptical rotation-time translation symmetry - $\hat{e}_{N,m}$**

$\hat{e}_{N,m}$ symmetry results in the following constraints:

$$\hat{L}^y{}_b \cdot \hat{R}^z{}_{N,m} \cdot \hat{L}^y{}_{1/b} \cdot \vec{F}_n = e^{in\frac{2\pi}{N}} \vec{F}_n \qquad (29)$$

The resulting selection rules dictate that the emission of $n = Nq \pm m$ (for integer $q$) harmonics is permitted with elliptical polarization and alternating helicity along the elliptical axis and within the plane orthogonal to the rotation axis, with the same underlying ellipticity $b$. The emission of $n = Nq$ harmonics is also permitted, but only with linear polarization parallel to the symmetry axis. This operation reduces to circular symmetry ($\hat{C}_{N,m}$) for $b = 1$.

**Elliptical improper rotation-time translation symmetry – even orders - $\hat{P}_{2N,m}$**

$\hat{P}_{2N,m}$ symmetry results in the following constraints:

$$[\hat{\sigma}_{xy} \cdot (\hat{L}^y{}_b \cdot \hat{R}^z{}_{2N,m} \cdot \hat{L}^y{}_{1/b})] \cdot \vec{F}_n = e^{in\frac{\pi}{N}} \vec{F}_n \qquad (30)$$

The resulting selection rules dictate that the emission of $n = 2Nq \pm m$ (for integer $q$) harmonics is permitted with elliptical polarization and alternating helicity along the elliptical axis and within the plane orthogonal to the rotation axis, with the same underlying ellipticity $b$. The emission of $n = N(2q + 1)$ harmonics is also permitted, but only with linear polarization parallel to the symmetry axis. This operation reduces to circular improper rotational symmetry ($\hat{M}_{2N,m}$) for $b = 1$.

**Elliptical improper rotation-time translation symmetry – odd orders - $\hat{P}_{2N+1,m}$**

$\hat{P}_{2N+1,m}$ symmetry results in the following constraints:

$$[\hat{\sigma}_{xy} \cdot (\hat{L}^y{}_b \cdot \hat{R}^z{}_{2N+1,m} \cdot \hat{L}^y{}_{1/b})] \cdot \vec{F}_n = e^{in\frac{\pi}{2N+1}} \vec{F}_n \qquad (31)$$

The resulting selection rules dictate that the emission of $n = 2q(2N + 1) \pm m$ (for integer $q$) harmonics is permitted with elliptical polarization and alternating helicity along the elliptical axis, and within the plane orthogonal to the rotation axis, with the same underlying ellipticity $b$. The emission of $n = (2N + 1)(2q + 1)$ harmonics is also permitted, but only with linear polarization parallel to the symmetry's rotation axis. This operation reduces to circular improper rotational symmetry ($\widehat{M}_{2N+1,m}$) for $b = 1$.

## S.4 Examples of dynamical groups in (2+1)D

Examples of dynamical groups in (2+1)D are given in Table S.1. In each group we specify the members of the group, and examples of vectorial functions which are characterized by the group.

**Table S.1| List of dynamical groups in (2+1)D, member DS operators, and example time-periodic vector functions that belong to each groups. A parametric Lissajou curve is given for each example field.**

| Members | Order | Example function: $\vec{E}(t) = \begin{pmatrix} E_x \\ E_y \end{pmatrix}$ | Example Lissajou curve |
|---|---|---|---|
| $\{\hat{1}\}$ | 1 | $\begin{pmatrix} \cos(\omega t) + \cos(2\omega t) + \sin(3\omega t) \\ \sin(\omega t) - \sin(2\omega t) + \cos(3\omega t) \end{pmatrix}$ | 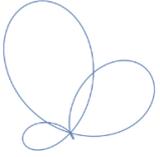 |
| $\{\hat{1}, \hat{C}_{n,m}, \dots, (\hat{C}_{n,m})^{n-1}\}$ | $n$ | $\begin{pmatrix} \cos(\omega t) + \cos(2\omega t) + \sin(4\omega t) \\ \sin(\omega t) - \sin(2\omega t) - \cos(4\omega t) \end{pmatrix}$ | 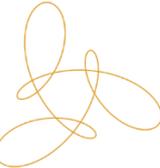 |
| $\{\hat{1}, \hat{e}_{n,m}, \dots, (\hat{e}_{n,m})^{n-1}\}$ | $n$ | $\begin{pmatrix} \cos(\omega t) + \cos(2\omega t) + \sin(4\omega t) \\ b\sin(\omega t) - b\sin(2\omega t) - b\cos(4\omega t) \end{pmatrix}$ | 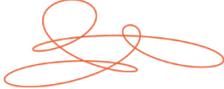 |
| $\{\hat{1}, \hat{T}\}$ | 2 | $\begin{pmatrix} \cos(3\omega t) \\ \cos(\omega t) + \cos(2\omega t) \end{pmatrix}$ | 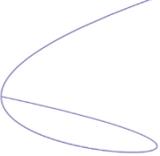 |
| $\{\hat{1}, \hat{S}\}$ | 2 | $\begin{pmatrix} \cos(\omega t) + \cos(2\omega t) \\ \sin(\omega t) + \sin(2\omega t) \end{pmatrix}$ | 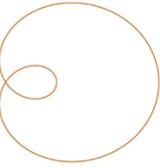 |
| $\{\hat{1}, \hat{Z}\}$ | 2 | $\begin{pmatrix} \cos(\omega t) \\ \sin(2\omega t + \pi/3) \end{pmatrix}$ | 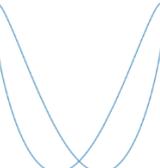 |
| $\{\hat{1}, \hat{H}\}$ | 2 | $\begin{pmatrix} \sin(\omega t) \\ \sin(2\omega t) + \cos(3\omega t)) \end{pmatrix}$ | 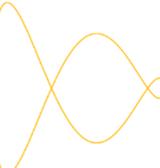 |
| $\{\hat{1}, \hat{Q}\}$ | 2 | $\begin{pmatrix} \sin(\omega t) \\ \sin(2\omega t) + \sin(3\omega t)) \end{pmatrix}$ | 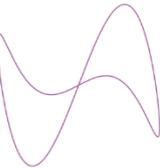 |
| $\{\hat{1}, \hat{G}\}$ | 2 | $\begin{pmatrix} \cos(\omega t) \\ \sin(2\omega t) + \cos(3\omega t)) \end{pmatrix}$ | 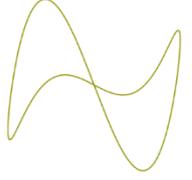 |
| $\{\hat{1}, \hat{D}_x, \hat{C}_2, \hat{H}_y\}$ | 4 | $\begin{pmatrix} \cos(\omega t) + \cos(3\omega t) \\ \sin(\omega t) + \sin(3\omega t)) \end{pmatrix}$ | 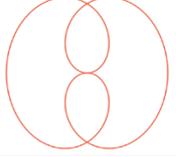 |
| $\{\hat{1}, \hat{H}_x, \hat{Q}, \hat{Z}_y\}$ | 4 | $\begin{pmatrix} \sin(\omega t) \\ \sin(2\omega t) \end{pmatrix}$ | 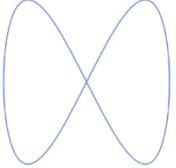 |
| $\{\hat{1}, \hat{Z}_x, \hat{G}, \hat{D}_y\}$ | 4 | $\begin{pmatrix} \cos(\omega t) - \cos(3\omega t) \\ \sin(2\omega t) \end{pmatrix}$ | 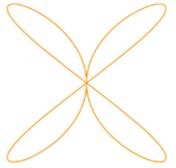 |
| $\{\hat{1}, \hat{C}_2, \hat{T}, \hat{G}\}$ | 4 | $\begin{pmatrix} \cos(\omega t) + \cos(3\omega t) \\ \cos(3\omega t) \end{pmatrix}$ | 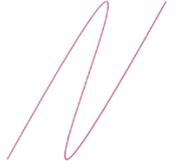 |
| $\{\hat{1}, \hat{Z}_x, \hat{T}, \hat{H}_x\}$ | 4 | $\begin{pmatrix} \cos(2\omega t) + \cos(4\omega t) \\ \cos(\omega t) \end{pmatrix}$ | 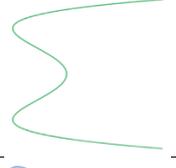 |
| $\{\hat{1}, \hat{C}_{n,m}, \dots, (\hat{C}_{n,m})^{n-1}, \hat{D}, \dots, \hat{D}(\hat{C}_{n,m})^{n-1}\}$ | $2n$ | $\begin{pmatrix} \cos(\omega t) + \cos(2\omega t) \\ \sin(\omega t) - \sin(2\omega t) \end{pmatrix}$ | 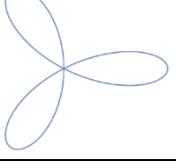 |
| $\{\hat{1}, \hat{e}_{n,m}, \dots, (\hat{e}_{n,m})^{n-1}, \hat{D}, \dots, \hat{D}(\hat{e}_{n,m})^{n-1}\}$ | $2n$ | $\begin{pmatrix} \cos(\omega t) + \cos(2\omega t) \\ b\sin(\omega t) - b\sin(2\omega t) \end{pmatrix}$ | 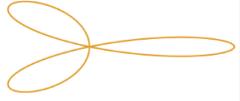 |

**S.5 Similarities with other groups**

As mentioned in the main text, dynamical Floquet groups are similar to those of line-groups[11] for the (2+1)D case. Considering line-groups that extend an infinite and periodic z-axis, this axis is similar to a periodic and infinite time-axis. Translations along the z-axis in line-groups are equivalent to time-translations ($\hat{\tau}_n$). All operations within the xy plane are identical to the spatial symmetry elements in (2+1)D dynamical groups. The main difference is in the time-reversal operation ($\hat{T}$), which geometrically is similar to a spatial reflection plane within the xy plane in line-groups. However, time-reversal is an anti-unitary operation unlike reflections. We further note that certain features of space-time groups[12], knot-groups[13], and space-groups[14], are isomorphic to Floquet groups. Due to the resemblence, we do not derive here the full list of character tables and multipcation tables of Floquet groups, but refer the reader to refs. 11–14, and references therein.

## S.6 TDSE calculations
### S.6.1 Numerical model
We numerically solve TDSE (2D/3D) in the length gauge, within the single active electron approximation, and the dipole approximation. In atomic units the TDSE is given by:

$$i\partial_t |\psi(t)\rangle = \left(-\frac{1}{2}\vec{\nabla}^2 + V_{atom}(\vec{r}) + \vec{r} \cdot \vec{E}(t)\right)|\psi(t)\rangle \tag{32}$$

where $|\psi(t)\rangle$ is the time dependent wave function of the single electron and $V_{atom}(\vec{r})$ is the atomic potential. We use a spherical atomic model, where the electron is initially in the 1s orbital ground state found by complex time propagation. The Coulomb interaction is softened at the origin with two free parameters set to describe the ionization potential ($I_p$) of the Ne atom ($I_p = 0.793\ a.u.$), and He$^+$ ion ($I_p = 1.999\ a.u.$). We set the atomic potential according to ref. 15:

$$V_{atom}(\vec{r}) = -\frac{z}{\sqrt{\vec{r}^2 + a}} \tag{33}$$

where $z = 1, 2$, and $a = 0.1195, 0.1583\ a.u.$, for Ne and He$^+$ in the 2D case, respectively, and z= 1.285, a=0.0001 for Ne in the 3D case. To avoid reflections from the boundaries, we use a complex absorbing potential set to:

$$V_{ab}(\vec{r}) = -i\eta(|\vec{r}| - r_0)^\alpha \Theta(|\vec{r}| - r_0) \tag{34}$$

where $\eta = 5 \times 10^{-4}$, $\alpha = 3$, and $r_0 = 36\ a.u.$ in the 2D case, and $\eta = 7 \times 10^{-4}$, $\alpha = 4$, and $r_0 = 32\ a.u.$ in the 3D case, and $\Theta$ is a Heaviside step function. The peak intensity of the laser ($I_0$) is set in the range of $10^{13}$-$10^{14}$ W/cm$^2$ in all calculations, such that the overall ionization does not exceed 5%. The driving pulse has a flat-top envelope, where the rise and drop sections are five optical cycles long, and the flat top is five optical cycles long in the 2D case, and the rise and drop sections are four optical cycles long, and the flat top is six optical cycles long in the 3D case, unless stated otherwise.

In order to solve Eq. (32), we used the 3$^{rd}$ order split operator method[16,17]. The time and spatial grids were discretized on an $L \times L$ Cartesian grid for $L = 120\ a.u.$ in 2D, and an $L \times L \times L$ Cartesian grid for $L = 90$ in 3D, with spacing $dx = dy = dz = 0.2348\ a.u.$, and $dt = 0.02\ a.u.$. Convergence was tested with respect to the grid densities, sizes, time step, and absorbing potential. The dipole acceleration was calculated using Ehrenfest theorem[18] as:

$$\vec{a}(t) = -\langle\psi(t)|\vec{\nabla}V_{atom} + \vec{E}(t)|\psi(t)\rangle \tag{35}$$

The harmonic spectra was calculated as the Fourier transform of Eq. (35). Projections of the circular components of the dipole acceleration are calculated according to:

$$\tilde{a}_{+/-}(\Omega) = \tilde{a}_x(\Omega) \pm i\tilde{a}_y(\Omega) \tag{36}$$

where (+/-) refers to harmonics with a positive/negative helicity, and "~" denotes the Fourier transformed component of the dipole.

### S.6.2 Numerical results
**(A) $\hat{Z}$ symmetry**

We numerically show that the (2+1)D reflection-translation DS, $\hat{Z} = \hat{\tau}_2 \cdot \hat{\sigma}$, results in the analytically derived selection rules. The results are presented in Figure S.1 for several exemplary cases. Each harmonic spectra shows the $x$ and $y$ components of the emitted harmonics with the numerically calculated ellipticity at each peak. The different drivers are presented in the inset of each harmonic spectra as a parametric Lissajous curve. The $x$-$y$ components of the field are given up to an overall amplitude coefficient. The analytically derived selection rules for drivers with $\hat{Z}$ symmetry are such that all harmonics are linearly-polarized, where even and odd harmonics are oriented along, and perpendicular, to the symmetry axis, respectively. As seen in Figure S.1, the harmonic selection rules are generally upheld very well by all harmonic orders, yet deviations can appear near resonances, which are marked with dashed lines. For instance, as shown in Fig. S.1(c), the ellipticity of the 10'th harmonic, which is near a resonance, is 0.09 instead of 0. As shown in Figure S.2, the deviation reduces when the length of the pulse or the length of the turn-on period of the flat-top envelope are increased. Thus, we conclude that resonances increase the sensitivity of HHG to broken time translation symmetry (e.g. finite pulses) – an important and potentially useful feature in HHG spectroscopy through DS breaking.

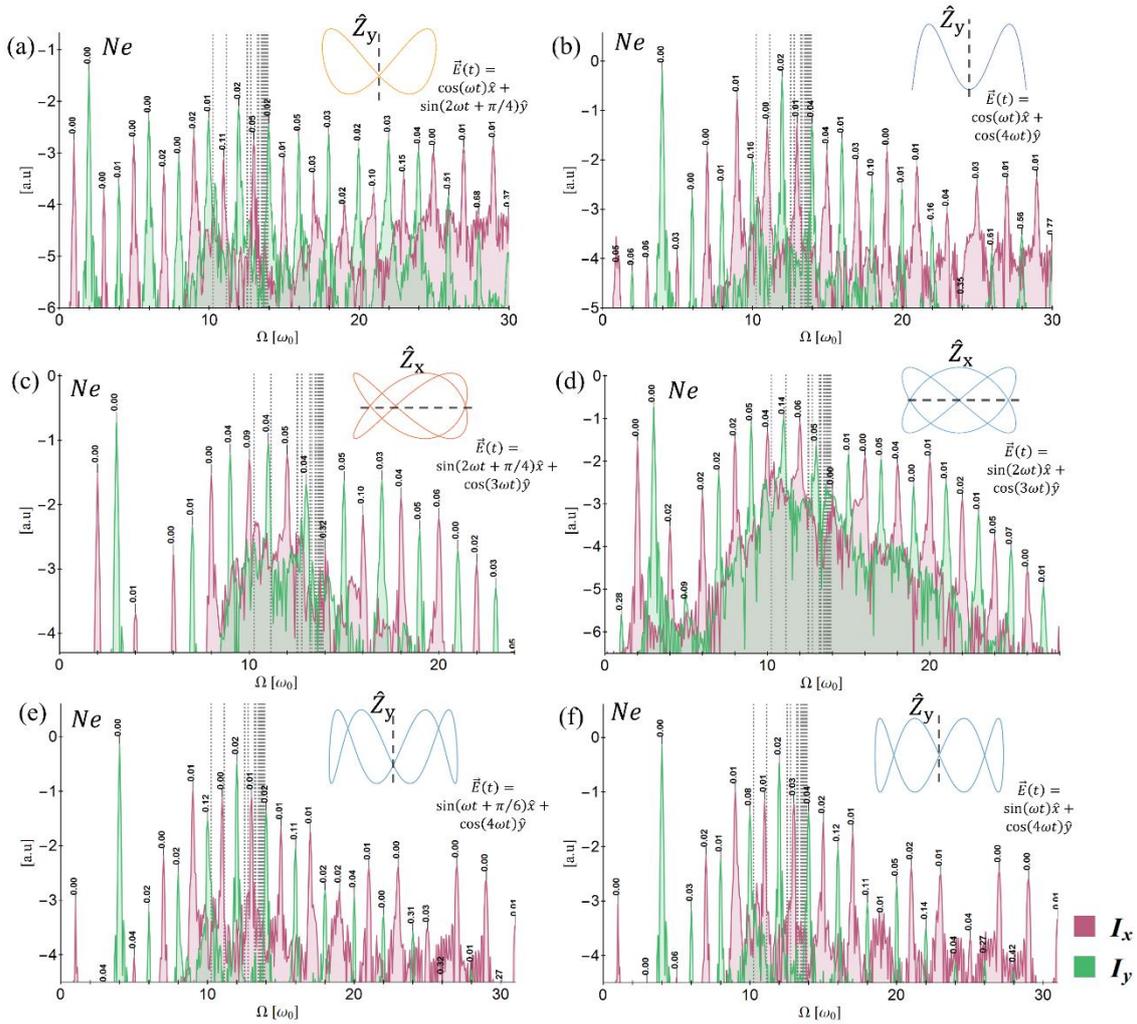

**Figure S.1| HG spectra for driver fields with $\hat{Z}$ DS from a model Ne atom system.** The symmetry is indicated in inset with the driver field up to an overall amplitude coefficient, and its Lissajous curve. Atomic resonances are indicated by dashed lines, where the last line indicates the $I_p$. According to the selection rules derived from group theory, all harmonics should be linearly polarized, even harmonics along the symmetry axis, and odd harmonics orthogonal to it. The calculated ellipticity is presented at each peak, indicated in black, and intensity in the y-axis is given in log scale.

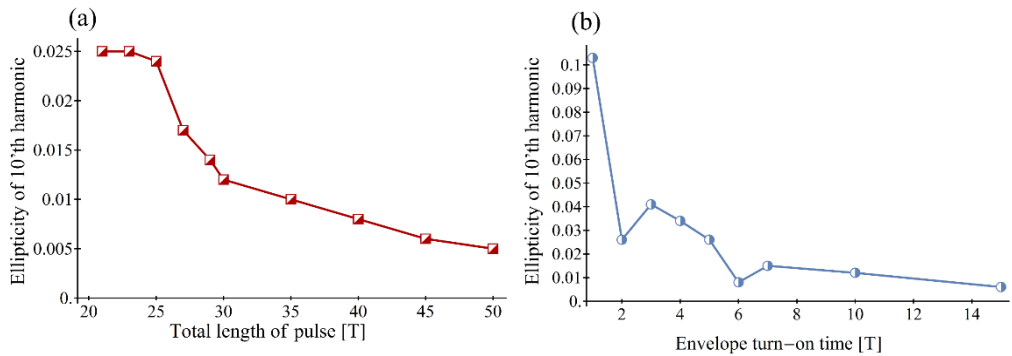

**Figure S.2| Time-translation symmetry breaking near atomic resonances.** The ellipticity of the 10'th harmonic near an atomic resonance from Fig S.1(c) is shown as a function of changes in the pulse envelope that perturb time-translation symmetry. **a** Ellipticity as a function of the total pulse duration for an envelope with a rise and drop sections 10 optical cycles long, and a flat top of varying duration. **b** Ellipticity as a function of varying turn-on duration in the flat-top envelope, with a flat-top part that is 10 optical cycles long. With increasing pulse duration (and turn-on duration), the ellipticity indeed approaches zero, as predicted by the selection rules for the symmetry in this system.

## (B) Time-reversal related symmetries ($\hat{Q}, \hat{G}, \hat{T}$)

Next, we examine (2+1)D DS operators that involve time-reversal symmetries ($\hat{T}, \hat{Q} = \hat{T} \cdot \hat{R}_2, \hat{G} = \hat{T} \cdot \hat{\tau}_2 \cdot \hat{R}_2$). According to group theory, these DSs should result in only linearly-polarized harmonics. However, these symmetries are slightly broken due to tunneling ionization (p. 9 in the main text). We numerically examine two systems with different ionization potentials: Ne and He$^+$. In He$^+$, the electron is tightly bound by the deep potential well, which results in a high $I_p$, and a large number of below $I_p$ harmonics that do not break time-reversal symmetry (since ionization does not play a role in their generation). Various examples are presented in Figures S.3 and S.4. Each harmonic spectra shows the harmonic intensity and the numerically calculated ellipticity at each peak. The different drivers are presented in the inset of each harmonic spectra as a parametric Lissajous curve. The $x$-$y$ components of the field are given up to an overall amplitude coefficient. As shown in Figure S.3 and S.4, the harmonic selection rules are generally upheld by harmonic orders below the $I_p$, where deviations are mostly found near resonances. Above $I_p$, the predicted selection rules are sometimes upheld and sometimes not. For example, in the spectrum of Figure S.3(b), several high harmonics have a relatively large ellipticity (up to 0.5) due to the broken symmetry. In contrast, in the spectrum of Figure S.4(b), all harmonics are practically linearly-polarized. This broken symmetry thus depends on several parameters, such as the shape and intensity of the driver and of the atomic potential.

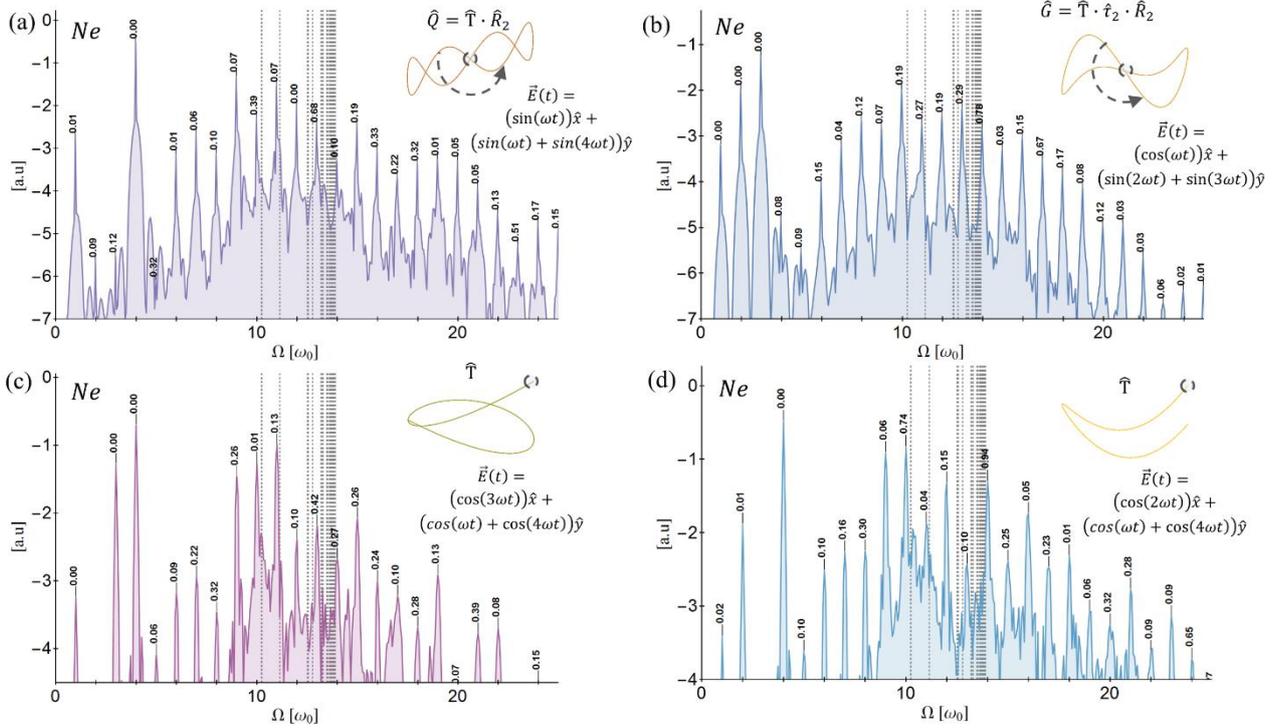

**Figure S.3| HG spectrum for driver fields with DSs that involve time-reversal from a model Ne atom system.** The symmetry is indicated in the inset of each spectrum with the driver field up to an overall amplitude coefficient, and its Lissajous curve. Atomic resonances are indicated by dashed lines, where the last line indicates the $I_p$. According to the selection rules derived from group theory all harmonics should be linearly-polarized. The calculated ellipticity is presented at each peak, indicated in black, and intensity in the y-axis is given in log scale.

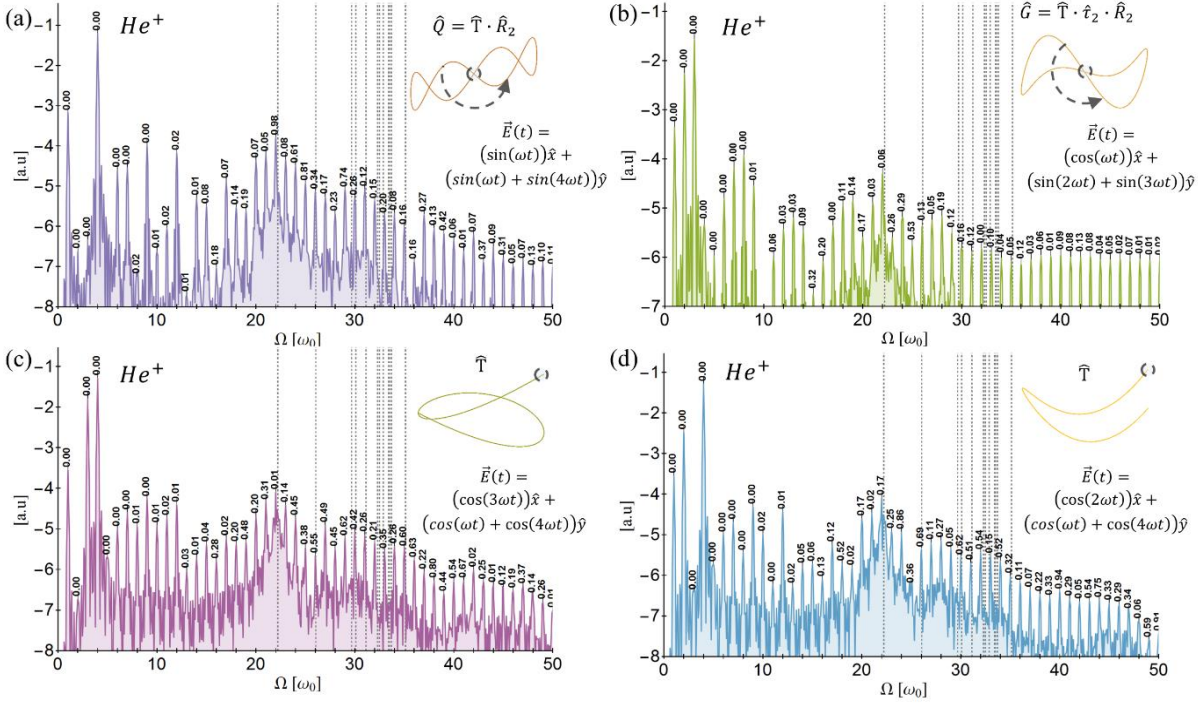

**Figure S.4| HG spectrum for driver fields with DSs involving time-reversal from a model He⁺ ion system.** The symmetry is indicated in inset of each spectrum with the driver field up to an overall amplitude coefficient, and its Lissajous curve. Atomic resonances are indicated by dashed lines, where the last line indicates the $I_p$. According to the selection rules derived from group theory all harmonics should be linearly-polarized. The calculated ellipticity is presented at each peak, indicated in black, and intensity in the y-axis is given in log scale.

## (C) 4$^{th}$ order elliptical symmetry ($\hat{e}_4$)

The kinetic energy and atomic/molecular potential terms are generally not invariant under the elliptical symmetry operation, $\hat{e}_n$. Still, there are cases where a driver pump can be chosen to uphold an elliptical symmetry, and not break the spherical symmetry of other terms in the Hamiltonian. This can only happen for n=4, since every ellipse has four points that are equidistant from the origin (Figure S.5). For example, in ref. 19 a train of linearly-polarized pulses was proposed that upholds this symmetry relation, and it was indeed shown numerically to yield the predicted selection rules. In this case, within each pulse the electron's motion, and therefore the polarization, is confined to the polarization axis of the pulse. Thus, the polarization is invariant under the elliptical DS. Below, we present an intriguing case that also yields the predicted selection rules of the (2+1)D elliptical DS, even though the electron's motion is two dimensional and intricate. We explore HHG driven by a synthetic field defined by a piecewise function that upholds 4'th order elliptical symmetry. The driver field has the form:

$$\vec{E}(t) = \frac{E_0}{2} A(t) \sin(2\omega_0 t) \times \begin{Bmatrix} \hat{x} + b\hat{y}; & 0 < t < T/4 \\ -\hat{x} + b\hat{y}; & T/4 < t < T/2 \\ -\hat{x} - b\hat{y}; & T/2 < t < 3T/4 \\ \hat{x} - b\hat{y}; & 3T/4 < t < T \end{Bmatrix} \quad (37)$$

where $E_0$ is the field amplitude, $A(t)$ is a flat-top envelope with a three optical cycle ramp up and down and a 15 optical cycle long flat-top, $b$ is the ellipticity of the underlying symmetry (from 0 to 1), and $\omega_0 = 2\pi/T$ is the optical frequency. Notably, producing this field requires an infinitely broad spectrum due to the non-smooth orientation shifts (the field is continuous but its derivative is not). Still, while such a field is not accessible experimentally, it can be approached by using multi-color drivers. This field provides 4$^{th}$ order elliptical symmetry present in all terms in the Hamiltonian. This example is slightly more general than the case of linearly-polarized pump trains, since the function in (37) maintains the symmetry on a sub-cycle level, where the electron dynamics is two dimensional and intricate. Numerical calculations of the HG spectra for several values of target ellipticity are shown in Figure S.6. Each peak shows the numerically calculated ellipticity, and the (+) and (-) projected helical components to indicate the helicity of the harmonics. The Lissajou parametric curve in each case is shown in inset, along with the ellipticity of the driver's underlying symmetry. In all cases the selection rules derived from group theory match with the calculated spectrum, except near atomic resonances.

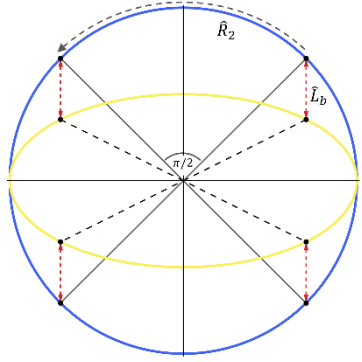

**Figure S.5| Schematic illustration of 4$^{th}$ order elliptical DS ($\hat{e}_4$) which still maintains circular DS for any ellipticity.** In every ellipse four points equidistant from the origin exist, which belong both to a circle and an ellipse simultaneously. The red arrows represent scaling transformation for the $y$ axis from the blue unit circle to the yellow ellipse.

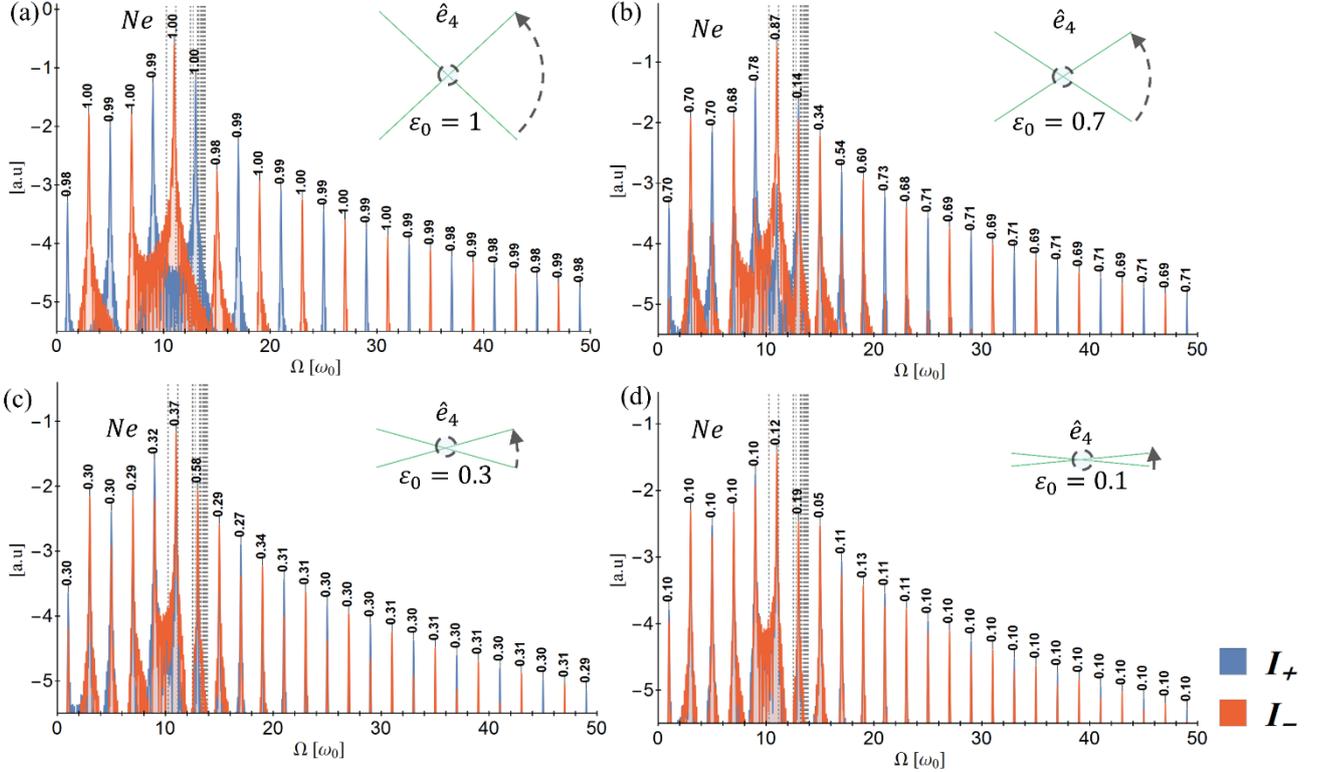

**Figure S.6| HG spectrum from a 4$^{th}$ order elliptical DS piecewise electric driver field, as in Eq. (37).** The ellipticity of the underlying symmetry is indicated in inset, as well as the field Lissajous curve. Calculated ellipticity for each harmonic peak is indicated in black, and according to the selection rules derived from group theory should be identical to the ellipticity of the driver's symmetry, with alternating helicities. Atomic resonances are indicated in dashed lines, where the last line indicates the $I_p$. Intensity in the y-axis is projected onto the left and right circularly polarized components, in red and blue, respectively, and the intensity is given in log scale.

### (D) (3+1)D Rotational symmetry in non-collinear HG ($\widehat{C}_{n,m}$)

Next, we examine (3+1)D rotational DS operators ($\hat{C}_{n,m} = \hat{\tau}_n \cdot \hat{R}_{n,m}$). These differ from the (2+1)D case by a specific choice of rotational symmetry axis, which we denote the $z$-axis. According to group theory, this DS should result in counter rotating $nq \pm m$ harmonics that are circularly polarized within the xy plane, and also $nq$ harmonics that can only be emitted with linear polarization along the z-axis (for integer $q$). Fascinatingly, emission of the $nq$ harmonics means that there is photon mixing between photons with all possible polarization axes (since otherwise the even $nq$ harmonics would be forbidden). We numerically verify this in a Ne like 3D system for a non-collinear geometry with $\hat{C}_3$ symmetry, where HG is driven by a counter rotating $\omega$-$2\omega$ bi-circular pump propagating along the z-axis, and an orthogonally propagating (along the x-axis) pulse of frequency $3\omega$ polarized linearly along the z-axis, seen in Figure S.7. The spectrum shows the harmonic intensity projected onto the circular components within the xy plane with the numerically calculated ellipticity at each peak, as well as the remaining z-axis emission. The driving field is presented in the inset as a parametric Lissajous curve, and schematic illustration of the non-collinear setup.

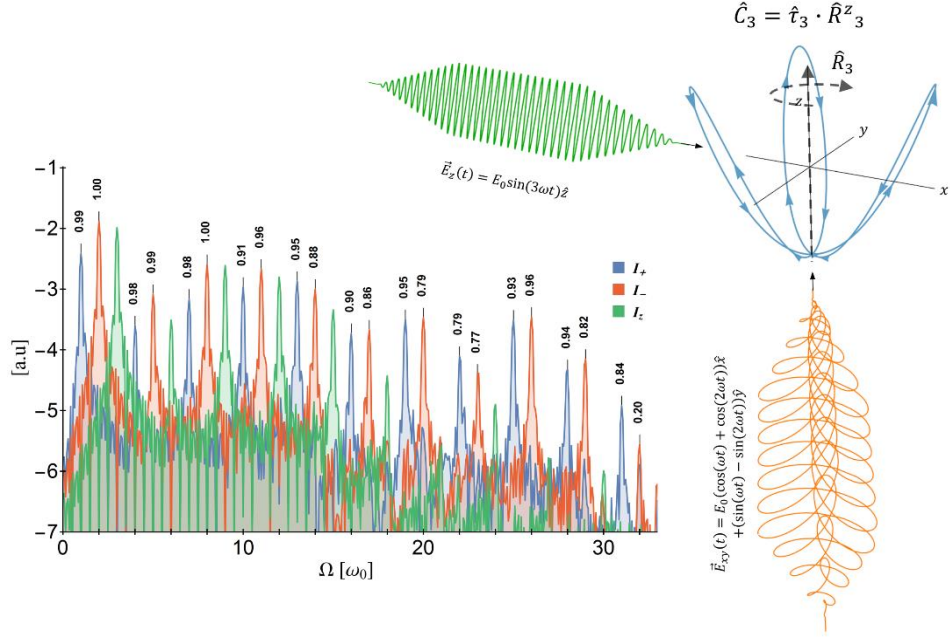

**Figure S.7| HG spectrum from a (3+1)D $\hat{C}_3$ symmetric pump.** The driving field is given in the inset up to an over amplitude coefficient. Red and blue stand for the circular projection within the xy plane, and calculated ellipticity for each harmonic peak within is indicated in black, and according to the selection rules should be exactly 1, with alternating helicities. The z-axis emission is presented in green, and is shown to be only at the predicted harmonic frequencies $3q$ (for integer $q$). Intensity is given in log scale.

**(E) (3+1)D Improper Rotational symmetry in noncollinear HG ($\hat{M}_{n,m}$)**

Next, we examine (3+1)D rotational DS operators ($\hat{M}_{2n,m} = \hat{\tau}_{2n} \cdot \hat{R}_{2n,m} \cdot \hat{\sigma}_h$). According to the presented theory, this DS results in counter rotating $2nq \pm m$ harmonics that are circularly polarized within the xy plane, and also $n(2q + 1)$ harmonics that can only be emitted with linear polarization along the z-axis (for integer $q$). The emission of the $n(2q)$ harmonics is symmetry forbidden, which is very surprising because this gives the appearance that selection rules from each axis are upheld separately, as if no mixing occurs. As an example, we explore $\hat{M}_{4,1}$ symmetry in a non-collinear geometry, with a bi-circular $\omega$-$3\omega$ counter-rotating pump pulse propagating along the z-axis, and an orthogonally propagating pump pulse (propagating along the x-axis) that is linearly polarized along the z-axis at frequency $2\omega$. In this case, harmonics 3,5,7,9,11… should be circularly polarized with alternating helicities within the $xy$ plane. Harmonics 2,6,10,14,18… should be emitted only with polarization components along the z-axis. The remaining 4,8,12,16… harmonics are symmetry forbidden. We numerically verify this in a Ne like system, seen in Figure S.8. The spectrum shows the harmonic intensity projected onto the circular components within the xy plane with the numerically calculated ellipticity at each peak, as well as the remaining z-axis emission. The driving field is presented in the inset as a parametric Lissajou curve, and schematic illustration of the non-collinear setup.

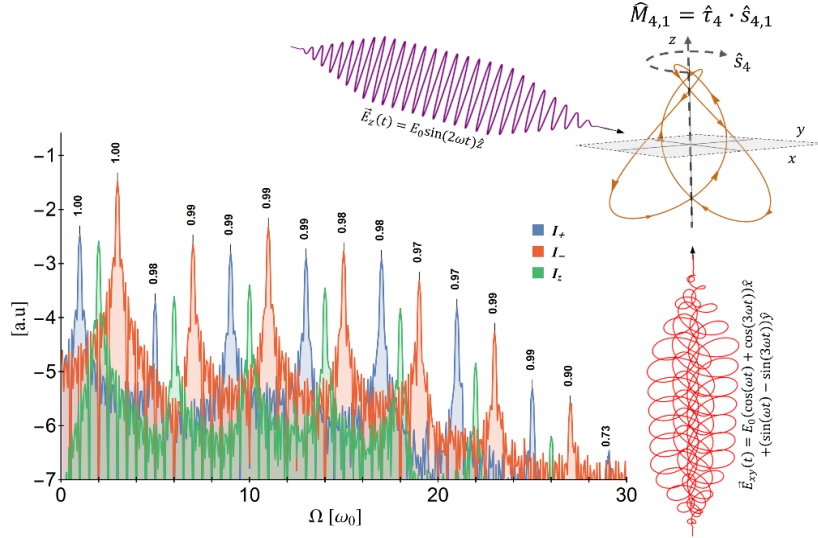

**Figure S.8| HG spectrum from a (3+1)D $\hat{M}_{4,1}$ symmetric pump.** The driving field is given in the inset up to an over amplitude coefficient. Red and blue stand for the circular projection within the xy plane, and calculated ellipticity for each harmonic peak within is indicated in black, and according to the selection rules should be exactly 1, with alternating helicities. The z-axis emission is presented in green, and is shown to be only at the predicted harmonic frequencies $2(2q+1)$ for integer $q$. Intensity is given in log scale.

## S.7 HG selection rules from molecules and molecular DS breaking spectroscopy

Here we present numerical examples for molecular HG selection rules, and for molecular symmetry and orientation spectroscopy from DS breaking. Specifically, we consider a planar 3-fold symmetric molecule that belongs to the $D_{3h}$ point group (which supports $120^0$ rotations, $180^0$ rotations, and also reflections along the major molecular axes, within the BO approximation), interacting with a pump field that exhibits $\hat{Z} = \hat{\tau}_2 \hat{\sigma}$ DS. We demonstrate that the selection rules derived from Floquet group theory depend both on the pump's DS, and on the symmetries of the molecule when it is oriented, and that the relative alignment with respect to the pump field determines whether a selection rule is observed (which is used for orientation spectroscopy). In case the medium is non-oriented, we show that the symmetry group of the orientation averaged ensemble and the pump's DS determine the observed selection rules.

According to the Floquet group theory, we should observe $\hat{Z}$ DS based selection rules in the HG spectrum from oriented molecules if the reflection axis of the pump is co-aligned with that of the molecules, leading to linearly polarized harmonics, where even/odd harmonics are polarized parallel/perpendicular to the symmetry axis. When the axes are not aligned, symmetry breaking will occur due to the molecular potential, and the selection rule will be broken. For the random non-oriented medium, theory predicts that the selection rule will be observed, because the orientation averaged ensemble is O(3) spherically symmetric (as discussed in the main text, and will only differ for chiral media), and supports the required reflection symmetry.

We consider the following 2D molecular potential that describes a poly-atomic molecule:

$$V_{mol}(x,y) = -\sum_{i=1}^{3} \frac{1/3}{\sqrt{(x-x_i)^2 + (y-y_i)^2}} \tag{38}$$

where we set $(x_i, y_i) = R_0(\cos(\alpha_i + \theta), \sin(\alpha_i + \theta))$, for $R_0 = 2.5\,a.u.$, $\alpha_i = \{0, 120^0, 240^0\}$, i=1,2,3, and we set $\theta$ as the relative angle of the molecule with respect to the $x$-axis. This potential is represented on a real-space Cartesian $L \times L$ grid for $L = 120$, with grid spacing $\Delta x = \Delta y = 0.1171$, where the ground state and first excited states are found by complex time propagation. The TDSE is solved as in section S.6 for the case where the molecule is interacting with the following cross-linear $\omega$-$2\omega$ driving field that possesses $\hat{Z}$ DS (i.e., dynamical reflection symmetry, but not 3-fold rotational DSs):

$$\vec{E}(t) = E_0 A(t)(\sin(\omega_0 t)\,\hat{x} + \sin(2\omega_0 t)\,\hat{y}) \tag{39}$$

where $A(t)$ is a flat-top envelope with a 4 optical cycle ramp up and down and an 6 optical cycle long flat-top, $\omega_0 = 2\pi/T$ is the optical frequency that corresponds to an 800nm wavelength, and $E_0$ is chosen such that the maximal intensity of the pulse is $10^{13}$[W/cm$^2$]. The TDSE is solved for many relative orientations with respect to the laser field, such that we obtain the HG spectrum as a function of $\theta$ and the HG spectrum from the random non-oriented media (see Figure S.9).

Figure S.9(a) shows that when the molecular axis is aligned with the pump's reflection axis, the selection rules for $\hat{Z}$ symmetry are observed. On the other hand, these selection rules are broken if the axes are not aligned (Figure S.9(b)). Furthermore, for the non-oriented medium, the selection rules from the pump's symmetry are upheld regardless of the molecular medium (Figure S.9(c)). In fact, most HG experiments deal with such non-oriented media, meaning usually the observed selection rules are identical to those observed from atomic media[20] (this is why the majority of numerical examples above are presented for the atomic case). We also note that the molecule's symmetry properties that are not exhibited by the pump field are 'washed out' in the spectrum, i.e., there is no selection rule due to 3-fold symmetry even if the medium is aligned. This is because a symmetry based selection rule is only observed if the full Hamiltonian supports the DS, and in this case the pump is not 3-fold symmetric. Importantly, similar results are obtained when the TDSE is solved with other initial states (and also from degenerate states, as long as an equal population of degenerate states is chosen), and in a full 3D molecular potential. All of these results match the predictions by the Floquet group theory.

Next, we consider DS breaking spectroscopy for the molecule with $\hat{Z}$ symmetry - we can identify the molecular orientation in aligned media by examining when the selection rule from the pump is 'restored' (orientation spectroscopy was demonstrated experimentally in SF$_6$ molecular media through a rotational DS in ref. 20). This is shown in Figure S.10(a) for the average intensity of the even/odd harmonic spectrum polarized along the x/y axes, respectively, and in Figure S.10(b) for the average ellipticity of the spectrum (harmonics 3-15). As seen, when the molecule is aligned with the pump, the selection rule is restored (hence $\theta$ can be measured), and we can also identify the presence of 3 reflection planes in the molecule. Conceptually, one can employ multiple pump fields of varying DSs and deduce the full molecular point group in this manner.

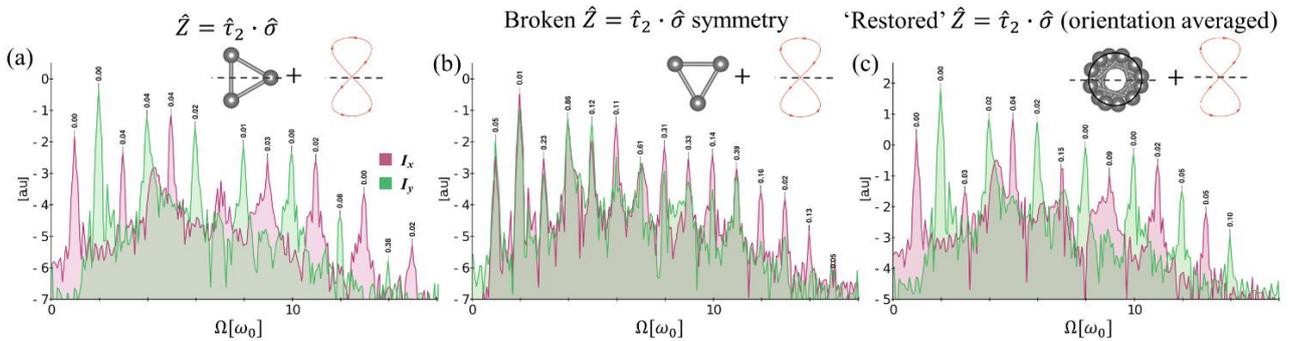

**Figure S.9| Molecular HG spectrum from a $\hat{Z}$ symmetric pump and 3-fold molecular potential.** The driving field's Lissajou plot is given in the inset together with the molecular orientation. The calculated ellipticity for each harmonic peak is indicated in black. Intensity is given in log scale. (a) Symmetric case, (b) symmetry broken case, (c) orientation averaged spectra (symmetric case).

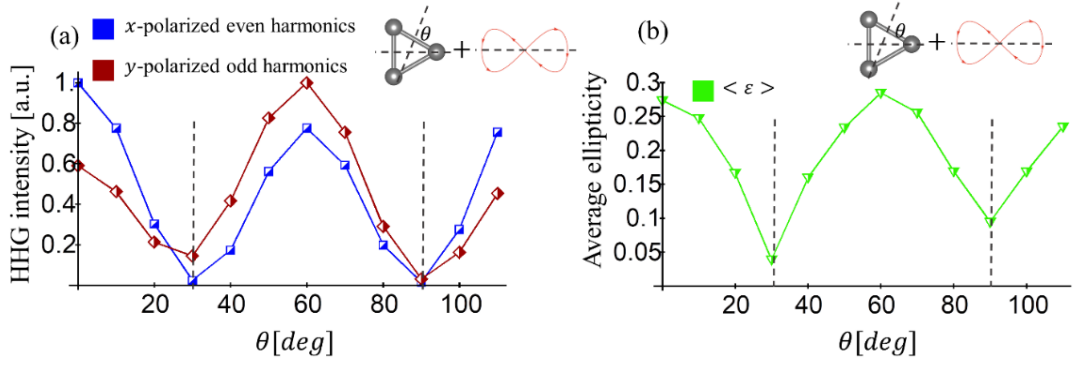

**Figure S.10| Molecular HG DS breaking orientation spectroscopy from a $\hat{Z}$ symmetric pump and a 3-fold molecular potential.** The driving field's Lissajou plot is given in the inset together with the molecule's orientation. (a) Intensity spectroscopy – the selection rules from the pump field dictate that x-polarized harmonics are only odd, and y-polarized harmonics are only even, which is upheld when the molecule's reflection axis is aligned with the pump. (b) Ellipticity spectroscopy – the selection rules from the pump field dictate that the spectrum is fully linearly polarized (0 ellipticity), which is upheld when the molecular axis is aligned with the pump.

## S.8 Examples for HG selection rules from solids and DS breaking orientation spectroscopy

Here we give some examples for new selection rules for HG from solids. In this particular case, the symmetries of the solid and the respective angle of incidence of the pump can determine the symmetries of the system. For example, a monochromatic laser field of any ellipticity upholds the DS $\hat{F} = \hat{\tau}_2 \cdot \hat{\iota}$. If the solid has an inversion center (centro-symmetry) $H_{HG}$ is invariant, and generation of even harmonics become forbidden. In fact, this is a well-known symmetry based selection rule in nonlinear optics[10,21] that is usually attributed to the symmetry of the solid, and our derivation presents its generalization. A second noteworthy example is that of the DS $\hat{Z} = \hat{\tau}_2 \cdot \hat{\sigma}$, which is a symmetry of any monochromatic linearly polarized laser field. When driving a solid with a reflection symmetry plane that is aligned orthogonal to the pump's polarization axis $H_{HG}$ is again invariant, leading to orthogonally linearly polarized even and odd harmonics. Importantly, if the solid is tilted out of plane with the incident wave, symmetry breaking occurs, which allows spectroscopically finding the solid's orientation. Lastly, DSs could be used to probe spin-orbit coupling or magnetic interactions in the solid, since these terms could break certain DSs.

As a numerical example, we show that the solid's orientation can break the reflection DS $\hat{Z}$ if it is not aligned to the pump's orientation, and use this to demonstrate orientation spectroscopy in solid HG. We calculate the harmonic spectrum from a single electron in a model 2D Mathieu type lattice potential[22]

$$V_{solid}(x,y) = -V_0 \left(1 + \cos\left(\frac{2\pi}{a}x\right)\right)\left(1 + \cos\left(\frac{2\pi}{a}y\right)\right) \quad (40)$$

where $V_0 = 0.25$, $a = 10$ [a.u]. This is done similarly to the atomic systems, on a real-space Cartesian grid spanning $20 \times 20$ lattice sites of size $L \times L$ for $L = 370$, with grid spacing $\Delta x = \Delta y = 0.2645$. The ground state is found by complex time propagation that corresponds to a real-space delocalized wave function that populates the first valence band, as is developed in ref. [23]. The TDSE is then solved as in section S.6 with the following cross-linear $\omega$-$2\omega$ driving field that possesses only $\hat{Z}$ DS:

$$\vec{E}(t) = E_0 A(t)(\sin(\omega_0 t + \phi)\hat{x} + \sin(2\omega_0 t)\hat{y}) \quad (41)$$

where $\phi = \pi/3$, $A(t)$ is a flat-top envelope with a 5 optical cycle ramp up and down and an 8 optical cycle long flat-top, $\omega_0 = 2\pi/T$ is the optical frequency that corresponds to an 800nm wavelength, and $E_0$ is chosen such that the maximal intensity of the pulse is $5 \times 10^{13}$[W/cm$^2$]. The harmonic spectrum is calculated and seen in Figure S.11 projected onto $x$ and $y$ polarization components for two cases: (a) the solid is aligned with the pump and the two share a y-axis reflection DS, and (b) the solid is rotated by $25^0$ with respect to the DS axis (this Mathieu potential supports three reflection axes). From Figure S.11, clearly when the DS is upheld the spectrum contains orthogonally polarized even and odd harmonics only, but for the broken symmetry different polarizations are emitted for different harmonics. We use this for orientation spectroscopy in Figure S.11(c), which shows the calculated average ellipticity of harmonics 10-18 in the spectrum as a function of the solid's angle with respect to the pump field. The average ellipticity has strong minima at $\alpha = 0^0, 45^0, 90^0$, which are the exact angles of reflection planes in the solid. Hence, an experimental ellipticity measurement may deduce the solid's orientation through DS breaking spectroscopy, and also the existence or lack of reflection planes in the solid. Notably we also see a peak in the average ellipticity at $\alpha \approx 25^0$, which is the orientation farthest away from any reflection axis, i.e., the maximal symmetry breaking may also be a useful observable.

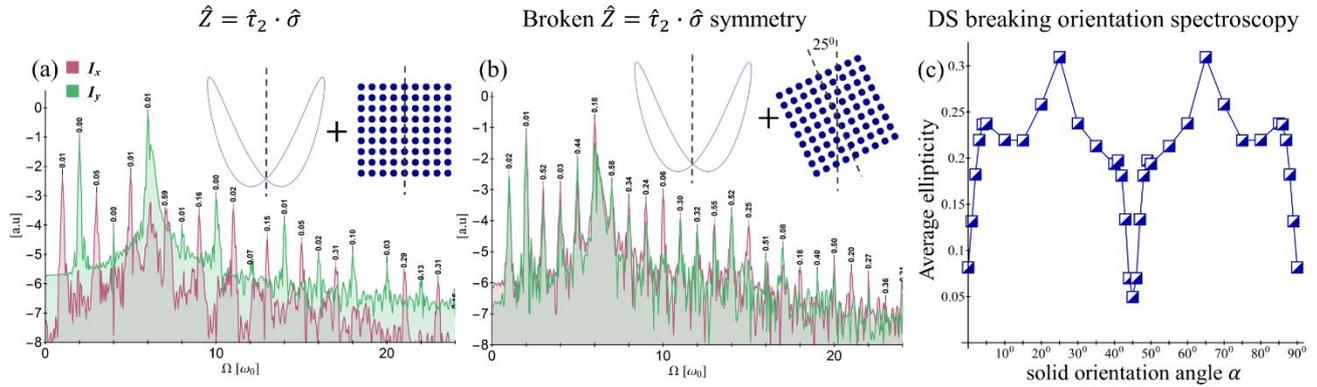

**Figure S.11| Solid HG spectrum from a $\hat{Z}$ symmetric pump and a Mathieu type potential used for orientation spectroscopy.** The driving field's Lissajou plot is given in the inset together with the solid's orientation. The calculated ellipticity for each harmonic peak is indicated in black. Intensity is given in log scale. (a) Symmetric case, (b) symmetry broken case, (c) average ellipticity of harmonics 10-18 in the spectrum with respect to the solid's orientation, where deep minima indicate reflection planes in the solid, and can be used to find the solid's orientation.

**References**

1. Kfir, O. *et al.* Generation of bright phase-matched circularly-polarized extreme ultraviolet high harmonics. *Nat. Photonics* **9,** 99–105 (2014).
2. Dudovich, N. *et al.* Measuring and controlling the birth of attosecond XUV pulses. *Nat. Phys.* **2,** 781–786 (2006).
3. Shafir, D., Mairesse, Y., Villeneuve, D. M., Corkum, P. B. & Dudovich, N. Atomic wavefunctions probed through strong-field light–matter interaction. *Nat. Phys.* **5,** 412–416 (2009).
4. Ben-Tal, N., Moiseyev, N. & Beswick, A. The effect of Hamiltonian symmetry on generation of odd and even harmonics. *J. Phys. B At. Mol. Opt. Phys.* **26,** 3017 (1993).
5. Weihe, F. A. *et al.* Polarization of high-intensity high-harmonic generation. *Phys. Rev. A* **51,** R3433–R3436 (1995).
6. Antoine, P., Carré, B., L'Huillier, A. & Lewenstein, M. Polarization of high-order harmonics. *Phys. Rev. A* **55,** 1314–1324 (1997).
7. Alon, O. E., Averbukh, V. & Moiseyev, N. Selection Rules for the High Harmonic Generation Spectra. *Phys. Rev. Lett.* **80,** 3743–3746 (1998).
8. Milošević, D. B. Circularly polarized high harmonics generated by a bicircular field from inert atomic gases in the p state: A tool for exploring chirality-sensitive processes. *Phys. Rev. A* **92,** 43827 (2015).
9. Fan, T. *et al.* Bright circularly polarized soft X-ray high harmonics for X-ray magnetic circular dichroism. *Proc. Natl. Acad. Sci. U. S. A.* **112,** 14206–11 (2015).
10. Yariv, A. & Yeh, P. *Photonics: optical electronics in modern communications*. **6,** (Oxford University Press New York, 2007).
11. Damnjanović, M. & Milošsević, I. in *Line Groups in Physics* (Springer, 2010).
12. Janssen, T., Janner, A. & Ascher, E. Crystallographic groups in space and time. *Physica* **41,** 541–565 (1969).
13. Neuwirth, L. P. *Knot Groups. Annals of Mathematics Studies.(AM-56)*. **56,** (Princeton University Press, 2016).
14. Hammond, C. & Hammond, C. *Basics of crystallography and diffraction*. **214,** (Oxford, 2001).
15. Medišauskas, L., Wragg, J., Van Der Hart, H. & Ivanov, M. Y. Generating Isolated Elliptically Polarized Attosecond Pulses Using Bichromatic Counterrotating Circularly Polarized Laser Fields. *Phys. Rev. Lett.* **115,** 153001 (2015).
16. Fleck, J. A., Morris, J. R. & Feit, M. D. Time-dependent propagation of high energy laser beams through the atmosphere. *Appl. Phys.* **10,** 129–160 (1976).

17. Feit, M. D., Fleck, J. A. & Steiger, A. Solution of the Schrödinger equation by a spectral method. *J. Comput. Phys.* **47,** 412–433 (1982).
18. Burnett, K., Reed, V. C., Cooper, J. & Knight, P. L. Calculation of the background emitted during high-harmonic generation. *Phys. Rev. A* **45,** 3347–3349 (1992).
19. Neufeld, O., Bordo, E., Fleischer, A. & Cohen, O. High Harmonics with Controllable Polarization by a Burst of Linearly-Polarized Driver Pulses. *Photonics* **4,** 31 (2017).
20. Baykusheva, D., Ahsan, M. S., Lin, N. & Wörner, H. J. Bicircular High-Harmonic Spectroscopy Reveals Dynamical Symmetries of Atoms and Molecules. *Phys. Rev. Lett.* **116,** 123001 (2016).
21. Tang, C. L. & Rabin, H. Selection Rules for Circularly Polarized Waves in Nonlinear Optics. *Phys. Rev. B* **3,** 4025–4034 (1971).
22. Slater, J. C. A soluble problem in energy bands. *Phys. Rev.* **87,** 807 (1952).
23. Liu, X. *et al.* Wavelength scaling of the cutoff energy in the solid high harmonic generation. *Opt. Express* **25,** 29216–29224 (2017).